\documentclass[aps,prc,superscriptaddress,twocolumn]{revtex4}

\usepackage{latexsym}
\usepackage{amssymb}
\usepackage{amsmath}
\usepackage{graphicx}
\usepackage{bm}
\usepackage{epsf}
\usepackage{epstopdf}
\usepackage{xcolor}

\newcommand{\nc}{\newcommand}       
\nc{\beq}       {\begin{eqnarray}}
\nc{\eeq}       {\end{eqnarray}}
\nc{\nn}        {\nonumber\\}
\nc{\mb}        {\mathbf}

\begin{document}

\title{Impact of strong magnetic fields on the inner crust of neutron stars}
\author{S. S. Bao}
\email{bao_shishao@163.com}
\affiliation{School of Physics and Information Engineering, Shanxi Normal University, Linfen 041004, China}
\author{J. N. Hu}
\email{hujinniu@nankai.edu.cn}
\affiliation{School of Physics, Nankai University, Tianjin 300071, China}
\author{H. Shen}
\email{shennankai@gmail.com}
\affiliation{School of Physics, Nankai University, Tianjin 300071, China}

\begin{abstract}
We study the impact of strong magnetic fields on the pasta phases that are expected to exist in the inner crust of neutron stars.
We employ the relativistic mean field model to describe the nucleon interaction
and use the self-consistent Thomas-Fermi approximation to calculate the nonuniform matter in neutron star crust.
The properties of pasta phases and crust-core transition are examined.
It is found that as the magnetic field strength $B$ is less than $10^{17}$ G,
the effects of magnetic field are not evident comparing with the results without magnetic field.
As $B$ is stronger than $10^{18}$ G,
the onset densities of pasta phases and crust-core transition density decrease significantly,
and the density distributions of nucleons and electrons are also changed obviously.
\end{abstract}

\maketitle



\section{Introduction}
\label{sec:1}

Neutron stars offer special natural laboratories for the study of nuclear physics
and astrophysics due to their extreme properties.
Neutron stars consist of extreme neutron-rich matter and their densities
can cover more than 10 orders of magnitude from surface to center~\cite{Lattimer04,Oertel17,Chamel08}.
It is generally believed that a neutron star mainly consists of four parts,
an outer crust of nuclei in a gas of electrons, an inner crust of neutron-rich nuclei with electron and neutron gas, a liquid outer core of homogeneous nuclear matter, and an inner core of exotic matter with non-nucleonic degrees of freedom~\cite{Chamel08,PR00,Webe05}.
From the neutron drip to the crust-core transition, i.e., the density range of inner crust,
the stable nuclear shape may change from droplet to rod, slab, tube, or bubble with increasing density.
As a result, the so-called nuclear pasta phases are expected to appear in the inner crust of neutron stars~\cite{Rave83,Mene08,Gril12,Okmaoto13},
which play a significant role in interpreting a lot of astrophysical observations,
such as the giant flares and quasiperiodic oscillations from soft $\gamma$-ray repeaters,
and glitches in the spin rate of pulsars~\cite{Delsate16,Stei08,Nandi16,Pons13,Thompson95,Wang13}.
The soft $\gamma$-ray repeaters and anomalous x-ray pulsars have already been confirmed as
magnetars with very strong surface magnetic fields~\cite{Usov92,Duncan92},
which can be as high as $10^{14}$-$10^{15}$ G~\cite{Rabhi15,web}.
The magnetic field strength in the core of a neutron star may even reach $10^{18}$ G~\cite{Muta19,Chatt15}.
So far, the mechanism and origin of strong magnetic fields in magnetars remain unclear,
and several hypotheses have been proposed
(see Ref.~\citep{Turo15} for a review and references therein).
Duncan and Thompson~\cite{Duncan92} suggested that such strong fields could be
generated by the dynamo mechanism in a rapidly rotating protoneutron star.
It has also been suggested that strong magnetic fields in neutron stars may result from
magnetic flux conservation during the collapse of a massive progenitor~\cite{Ferr06}.
It is still under discussion how strong the magnetic fields can be in the crust and interior
of neutron stars.

In past decades, great efforts have been devoted to study
the effects of strong magnetic fields on the properties of
asymmetry nuclear matter and neutron star structures,
and the homogeneous stellar matter under strong magnetic fields has also been extensively studied~\cite{Rabhi15,Broderick00,Yue06,Dong13,Aguirre11,Garc11}.
The effects of Landau quantization can reduce the electron chemical potential and increase the proton fraction,
which leads to the softening of equation of state for neutron stars.
The hyperonic matter appearing in the core of neutron stars under strong magnetic fields were studied in Ref.~\cite{Yue09}, where it was found that the onset densities of hyperons could be observably changed by strong magnetic fields.
Furthermore, magnetization and magnetic susceptibility properties of cold neutron star matter and even
the warm stellar matter were also examined within different methods~\cite{Dong13,Rabhi11,Aguirre14}.
However, the studies on nonuniform crust matter under strong magnetic fields are rare due to
the complex structures of pasta phases.
Recently, some researchers studied
the density ranges and proton fractions in neutron star crusts under strong magnetic fields
by analyzing the dynamical instability region of ``$npe$" matter with various models~\cite{Avancini18,Chen17,Fang16,Fang17}.
The neutron drip densities with strong magnetic fields were calculated using
Brussels-Montreal microscopic nuclear mass models in Ref.~\cite{Fantina16}.
The magnetic susceptibility and electron transport properties in the neutron star crust
with strong magnetic fields were reported in Refs.~\cite{Muta19,Bland82,Yakovlev15}.
However, most studies do not take into account the nuclear pasta structures in the inner crust of neutron stars.
In Ref.~\cite{Lima13}, the nuclear pasta phases were studied using
the relativistic mean field (RMF) models with NL3~\cite{NL3} and TM1~\cite{TM1} parametrizations under strong magnetic fields $\approx$$10^{17}$-$10^{18}$ G,
where the proton fraction was fixed and the anomalous magnetic moments of nucleons were neglected.
The $npe$ matter satisfying $\beta$ equilibrium condition was studied using SkM nucleon-nucleon interaction
in Ref.~\cite{Nandi11}, where only the droplet phase was considered.
Therefore, it is interesting and important to perform further investigations
on the nonuniform matter in the inner crust of neutron stars under magnetic fields.

In order to evaluate the influence of magnetic fields, the field strength at a given location
in the star must be known. However, it is generally believed that the magnetic field
configuration in a neutron star is very complex and difficult to determine~\cite{Turo15,Chatterjee19}.
Only the surface magnetic field can be obtained from related astrophysical observations,
whereas the internal magnetic field of the star cannot be directly accessible to observations.
Due to the complexity in dealing with Maxwell's equations,
a number of parameterized models have been proposed to describe the magnetic field distribution
in neutron stars~\cite{Chatterjee19,Band97,Ferr10,Dexheimer12,Lopes15,Dexheimer17}.
In Ref.~\citep{Chatterjee19}, the authors presented a magnetic field profile from the surface
to the interior of the star, where the magnetic field strength corresponding to the inner crust
area could be as large as $\approx 10^{17}$ G for a central field strength of $5 \times 10^{17}$ G.
In the present work, we focus on the effects of strong magnetic fields in the inner crust
with a thickness of less than 1 km. For simplicity, we neglect the variation of the field strength
within this narrow range of radial distance and assume a homogeneous magnetic field along the $z$ direction.

We employ the Wigner-Seitz (WS) approximation to describe the inner crust and
use the self-consistent Thomas-Fermi (TF) approximation to calculate the nonuniform matter
with considering various pasta configurations.
In the TF approximation,
the surface energy and the distributions of nucleons and electrons are treated self-consistently.
We adopt the RMF model to describe nucleon-nucleon interaction.
In the RMF model, nucleons interact with each other via the exchange of scalar and vector mesons.
We use two different RMF parametrizations, TM1 and IUFSU~\cite{IUFSU},
which are successful in describing the ground-state properties of finite nuclei and compatible with maximum neutron-star mass $\approx$$2 M_\odot$.
The TM1 model has been successfully used to
construct the equation of state for neutrons stars and supernova simulations~\cite{Shen11}.
Compared with TM1 model,
an additional $\omega$-$\rho$ coupling term is added in IUFSU model,
which plays an important role in
modifying the density dependence of symmetry energy and affects the neutron star properties~\cite{IUFSU,Bao15}.
The symmetry energy slope $L$ in TM1 model is as large as $110.8$ MeV,
while $L$ in IUFSU model is $40.7$ MeV.
By comparing the results from these two models,
it is helpful for understanding the impacts of nuclear symmetry energy on pasta phases with strong magnetic fields.

This paper is organized as follows.
In Sec.~\ref{sec:2},
we briefly describe the RMF model and present the formalism used in this study.
In Sec.~\ref{sec:3},
We show the numerical results and discuss the influence of strong magnetic fields on the
properties of pasta phases and the crust-core transition of neutron star.
Section~\ref{sec:4} is devoted to the conclusions.

\section{Formalism}
\label{sec:2}

We employ the TF approximation to study the inner crust of neutron stars with strong magnetic fields.
The nucleon interaction is described by the RMF model, where the nucleons interact
through the exchange of various mesons,
and the charged particles interact through electromagnetic field $A^{\mu}$.
The isoscalar-scalar meson $\sigma$,
isoscalar-vector meson $\omega$, and isovector-vector meson $\rho$ are taken
into account. For a system consisting of protons, neutrons, and electrons,
the Lagrangian density is given by
\begin{eqnarray}
\label{eq:LRMF}
\mathcal{L}_{\rm{RMF}} & = & \sum_{i=p,n}\bar{\psi}_i
\left\{i\gamma_{\mu}\partial^{\mu}-{\left(M+g_{\sigma}\sigma\right)}
-\frac{1}{2}{\kappa_i}{\sigma_{\mu\nu}}{F^{\mu\nu}} \right.  \notag \\
& & \left. -\gamma_{\mu} \left[g_{\omega}\omega^{\mu} +\frac{g_{\rho}}{2}\tau_a\rho^{a\mu}
+\frac{e}{2}\left(1+\tau_3\right)A^{\mu}\right]\right\}\psi_i  \notag  \\
& & +\bar{\psi}_{e}\left[i\gamma_{\mu}\partial^{\mu} -m_{e} +e \gamma_{\mu}
A^{\mu} \right]\psi_{e}  \notag \\
&& +\frac{1}{2}\partial_{\mu}\sigma\partial^{\mu}\sigma -\frac{1}{2}%
m^2_{\sigma}\sigma^2-\frac{1}{3}g_{2}\sigma^{3} -\frac{1}{4}g_{3}\sigma^{4}
\notag \\
&& -\frac{1}{4}W_{\mu\nu}W^{\mu\nu} +\frac{1}{2}m^2_{\omega}\omega_{\mu}%
\omega^{\mu} +\frac{1}{4}c_{3}\left(\omega_{\mu}\omega^{\mu}\right)^2  \notag
\\
&& -\frac{1}{4}R^a_{\mu\nu}R^{a\mu\nu} +\frac{1}{2}m^2_{\rho}\rho^a_{\mu}%
\rho^{a\mu} \notag \\
&& +\Lambda_{\rm{v}} \left(g_{\omega}^2
\omega_{\mu}\omega^{\mu}\right)
\left(g_{\rho}^2\rho^a_{\mu}\rho^{a\mu}\right) -\frac{1}{4}%
F_{\mu\nu}F^{\mu\nu},
\end{eqnarray}
where $W^{\mu\nu}$, $R^{a\mu\nu}$, and $F^{\mu\nu}$ are the antisymmetric field tensors corresponding to $\omega^{\mu}$, $\rho^{a\mu}$, and $A^{\mu}$, respectively.
$\kappa_i$ ($i=p,n$) denotes the anomalous magnetic moment of nucleons.
In the RMF approximation, the meson fields are treated as classical fields,
and the field operators are replaced by their expectation values.
For a static system, the nonvanishing expectation values are
$\sigma =\left\langle \sigma \right\rangle$,
$\omega =\left\langle\omega^{0}\right\rangle$,
$\rho =\left\langle \rho^{30} \right\rangle$,
and $A =\left\langle A^{0}\right\rangle$.
From the Lagrangian density (\ref{eq:LRMF}), we can obtain the equations of motion for meson fields and electromagnetic field,
\begin{eqnarray}
&&-\nabla ^{2}\sigma +m_{\sigma }^{2}\sigma +g_{2}\sigma ^{2}+g_{3}\sigma
^{3}=-g_{\sigma }\left( n_{p}^{s}+n_{n}^{s}\right) ,
\label{eq:eqms} \\
&&-\nabla ^{2}\omega +m_{\omega }^{2}\omega +c_{3}\omega^{3}
+2\Lambda_{\rm{v}}g^2_{\omega}g^2_{\rho}{\rho}^2 \omega
=g_{\omega}\left( n_{p}+n_{n}\right) ,
\label{eq:eqmw} \\
&&-\nabla ^{2}\rho +m_{\rho }^{2}{\rho}
+2\Lambda_{\rm{v}}g^2_{\omega}g^2_{\rho}{\omega}^2{\rho}
=\frac{g_{\rho }}{2}\left(n_{p}-n_{n}\right) ,
\label{eq:eqmr} \\
&&-\nabla ^{2}A=e\left( n_{p}-n_{e}\right) ,
\label{eq:eqma}
\end{eqnarray}
where $n_i^s$ and $n_i$ represent the scalar and vector densities of nucleons, respectively.

For a nonuniform nuclear system at zero temperature, the local energy density including Coulomb energy is given by
\begin{eqnarray}
{\varepsilon }_{\rm{rmf}}(r) &=&\displaystyle{\sum_{i=p,n,e}}{\varepsilon_i}
+g_{\omega }\omega \left( n_{p}+n_{n}\right)
+\frac{g_{\rho }}{2}\rho \left( n_{p}-n_{n}\right)
\notag \\
&&+\frac{1}{2}(\nabla \sigma )^{2}+\frac{1}{2}m_{\sigma }^{2}\sigma ^{2}+%
\frac{1}{3}g_{2}\sigma ^{3}+\frac{1}{4}g_{3}\sigma ^{4}  \notag \\
&&-\frac{1}{2}(\nabla \omega )^{2}-\frac{1}{2}m_{\omega }^{2}\omega ^{2}-%
\frac{1}{4}c_{3}\omega ^{4}
\notag \\
&&-\frac{1}{2}(\nabla \rho )^{2}-\frac{1}{2}m_{\rho }^{2}\rho ^{2}
-\Lambda_{\rm{v}}g_{\omega }^{2}g_{\rho }^{2}\omega ^{2}\rho ^{2} \notag  \\
&&-\frac{1}{2}(\nabla A)^{2}+eA\left( n_{p}-n_{e}\right).
\label{eq:ETF}
\end{eqnarray}
In order to study the effects of strong magnetic fields on neutron star crust,
we assume that the nuclear system is in an external homogeneous magnetic field
$\mathbf{B}$ along the $z$ direction, $A^{\mu}=(0,0,Bx,0)$.
So the proton scalar density $n_{p}^{s}$ and proton vector density $n_p$ are given by
\begin{eqnarray}
n_{p}^{s} &=&\frac{{eB}M^{\ast }}{2\pi ^{2}}\sum_{\nu }\sum_{s}
\left(\frac{\sqrt{M^{\ast 2}+2\nu {eB}}-s\kappa _{p}B}{\sqrt{M^{\ast
2}+2\nu {eB}}} \right.  \notag  \\
&& \left.\times \ln \left\vert \frac{k_{F,\nu ,s}^{p}+E_{F}^{p}}{\sqrt{
M^{\ast 2}+2\nu {eB}}-s\kappa_{p}B}\right\vert \right),
\label{eq:nps}
\end{eqnarray}
\begin{eqnarray}
n_{p}=\frac{{e}B}{2\pi ^{2}}\sum_{\nu }\sum_{s}k_{F,\nu,s}^{p},
\label{eq:np}
\end{eqnarray}
and the proton energy density $\varepsilon_p$ in Eq.~(\ref{eq:ETF}) is written as
\begin{eqnarray}
{\varepsilon_p} &=& \frac{eB}{4{\pi}^2}{\sum_{\nu}}{\sum_{s}}
\left[{k_{F,\nu,s}^{p}}{E_{F}^{p}}+
{\left(\sqrt{{M^{\ast}}^{2}+2{\nu}eB}-s{\kappa_p}B\right)^2} \right.  \notag  \\
&& \left.\times{\ln\left|\frac{{k_{F,\nu,s}^{p}}+{E_{F}^{p}}}{\sqrt{{M^{\ast}}^{2}+2{\nu}eB}-s{\kappa_p}B} \right|}\right],
\label{eq:ep}
\end{eqnarray}
where $k_{F,\nu,s}^{p}$ is the Fermi momentum of proton with spin $s$ and Landau level $\nu$,
and $M^{\ast}=M+{g_{\sigma}}{\sigma}$ is the effective nucleon mass.
The Fermi energy of proton is given by
\begin{eqnarray}
{E_{F}^{p}}=\sqrt{{k_{F,\nu,s}^{p2}}+{\left(\sqrt{{M^{\ast}}^2+2{\nu}eB}-s{\kappa_p}B\right)}^2}
\label{eq:epf}.
\end{eqnarray}
We notice that $\nu=0, 1, 2, \ldots, \nu_{\rm{max}}$,
\begin{eqnarray}
\nu_{\rm{max}}=\left[\frac{\left({{E_{F}^{p}}+s \kappa_p B}\right)^2-{M^{\ast}}^{2}}{2eB}\right],
\end{eqnarray}
where [$x$] means the largest integer which is not larger than $x$.
The neutron scalar density $n_{n}^{s}$ and neutron vector density $n_n$ are given by
\begin{eqnarray}
n_{n}^{s} &=&\frac{M^{\ast }}{4\pi ^{2}}\sum_{s}\left[
k_{F,s}^{n}E_{F}^{n}-\left( M^{\ast }-s{\kappa _{n}B}\right)^{2}  \right.  \notag  \\
&& \left.\times\ln\left\vert \frac{k_{F,s}^{n}+E_{F}^{n}}{M^{\ast }-s{\kappa _{n}B}}\right\vert \right],
\label{eq:nns}
\end{eqnarray}
\begin{eqnarray}
n_{n} &=& \frac{1}{2\pi ^{2}}\sum_{s}\left\{ \frac{1}{3}k_{F,s}^{n3}-
\frac{1}{2}s\kappa _{n}B\left[ \left( M^{\ast }-s{\kappa _{n}B}\right)
k_{F,s}^{n} \right. \right. \nonumber \\
 & & \left. \left. +E_{F}^{n2}\left( \arcsin \frac{M^{\ast }-s{\kappa _{n}B}}
{E_{F}^{n}}-\frac{\pi }{2}\right) \right] \right\} ,
\label{eq:nn}
\end{eqnarray}
and the neutron energy density $\varepsilon_n$ in Eq.~(\ref{eq:ETF}) is written as
\begin{eqnarray}
{\varepsilon_n} &=& \frac{1}{4{\pi}^2}{\sum_{s}} \left\{
\frac{1}{2}{k_{F,s}^{n}}{E_F^n}^3   \right.  \notag  \\
&& -\frac{2}{3}s{\kappa_n}B{{E_F^n}^3}
\left(\arcsin{\frac{M^{\ast}-s{\kappa_n}B}{E_F^n}}-\frac{\pi}{2}\right)  \notag \\
&&-\left(\frac{s{\kappa_n}B}{3}+\frac{M^{\ast}-s{\kappa_n}B}{4}\right) \notag \\
&&\times \left. \left[\left(M^{\ast}-s{\kappa_n}B\right){k_{F,s}^n}{E_F^n} \right. \right. \notag \\
&& +\left. \left. \left(M^{\ast}-s{\kappa_n}B\right)^3
\ln\left|{\frac{{k_{F,s}^n}+{E_F^n}}{M^{\ast}-s{\kappa_n}B}}\right|\right]\right\}
\label{eq:en},
\end{eqnarray}
where $k_{F,s}^{n}$ is the Fermi momentum of neutron with spin $s$.
The Fermi energy of neutron is given by
\begin{eqnarray}
{E_{F}^{n}}=\sqrt{{k_{F,s}^{n2}}+\left(M^{\ast}-s{\kappa_n}B\right)^2}.
\label{eq:enf}
\end{eqnarray}
The electron density is given by
\begin{equation}
n_e=\frac{eB}{2\pi ^{2}}\sum_{\nu
}\sum_{s}k_{F,\nu ,s}^{e},
\end{equation}
and the electron energy density $\varepsilon_e$ in Eq.~(\ref{eq:ETF}) is written as
\begin{eqnarray}
{\varepsilon_e} &=& \frac{eB}{4{\pi}^2}{\sum_{\nu}}{\sum_{s}}
\left[{k_{F,\nu,s}^{e}}{E_{F}^{e}}+\left(m_e^2+2{\nu}eB\right)  \right. \notag  \\
&& \left.\times\ln\left|\frac{{k_{F,\nu,s}^{e}}+{E_{F}^{e}}}{\sqrt{m_e^{2}+2{\nu}eB}}\right|\right],
\end{eqnarray}
where $k_{F,\nu,s}^{e}$ is the Fermi momentum of electron with spin $s$ and Landau level $\nu$,
and the Fermi energy of electron is given by
\begin{eqnarray}
{E_{F}^{e}}=\sqrt{{k_{F,\nu,s}^{e2}}+{m_e^2}+2{\nu}eB}.
\end{eqnarray}
For simplicity, the anomalous magnetic moment of electron is neglected in our calculation.
So, the largest Landau level $\nu_{\rm{max}}$ of electron is given by
\begin{eqnarray}
\nu_{\rm{max}}=\left[\frac{{E_{F}^{e}}^2-m_e^2}{2eB}\right],
\end{eqnarray}
where the meaning of [$x$] is the same as the case of protons.
We should point out that the energy density from the contribution of electromagnetic field, $B^2/8{\pi}^2$,
is neglected in our calculation,
which does not affect the phase transitions of different pasta phases and crust-core transitions.

\begin{table*}[htb]
\caption{Parameter sets used in this work. The masses are given in MeV.}
\begin{center}
\begin{tabular}{lccccccccccc}
\hline\hline
Model   &$M$  &$m_{\sigma}$  &$m_\omega$  &$m_\rho$  &$g_\sigma$  &$g_\omega$
        &$g_\rho$ &$g_{2}$ (fm$^{-1}$) &$g_{3}$ &$c_{3}$ &$\Lambda_{\textrm{v}}$ \\
\hline
TM1     &938.0  &511.198  &783.0  &770.0  &10.0289  &12.6139  &9.2644
        &$-$7.2325   &0.6183   &71.3075   &0.000  \\
IUFSU   &939.0  &491.500  &782.5  &763.0  &9.9713   &13.0321  &13.5900
        &$-$8.4929   &0.4877   &144.2195  &0.046 \\
\hline\hline
\end{tabular}
\label{tab:1}
\end{center}
\end{table*}
\begin{table}[htb]
\caption{Saturation properties of nuclear matter for the TM1 and IUFSU models.
The quantities $E_0$, $K$, $E_{\text{sym}}$, and $L$ are, respectively,
the energy per nucleon, incompressibility coefficient, symmetry
energy, and symmetry energy slope at saturation density $n_0$.}
\label{tab:2}
\begin{center}
\begin{tabular}{l c c c c c}
\hline\hline
Model & $n_0$ (fm$^{-3}$) & $E_0$ (MeV) & $K$ (MeV) & $E_{\text{sym}}$ (MeV) & $L$ (MeV) \\
\hline
TM1   & 0.145 & $-$16.3 & 281.0 & 36.9 & 110.8 \\
IUFSU & 0.155 & $-$16.4 & 231.0 & 31.3 & 47.2  \\
\hline\hline
\end{tabular}
\end{center}
\end{table}
\begin{figure}[htb]
\includegraphics[bb=10 38 535 786, width=7 cm,clip]{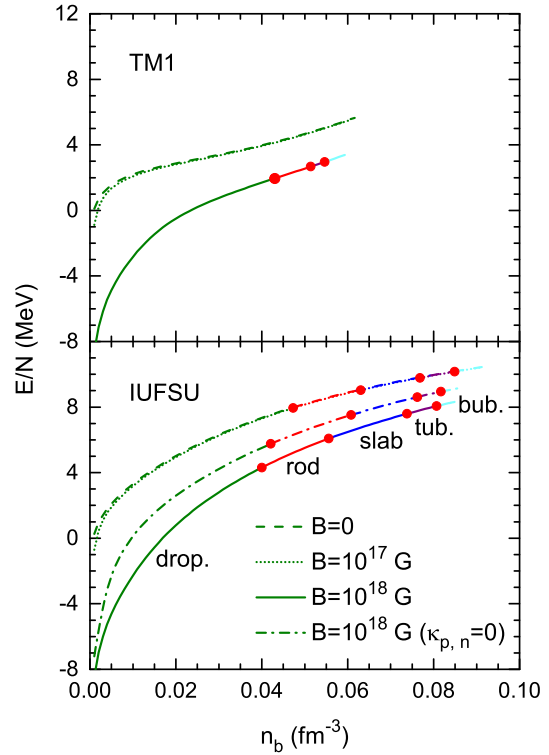}
\caption{(Color online) Binding energy per nucleon $E/N$ of pasta phases as a function of
baryon density $n_b$ for TM1 (upper panel) and IUFSU (lower panel) models
with different magnetic field strength,
$B=0$ (dashed line), $B=10^{17}$ G (dotted line), and $B=10^{18}$ G (solid line).
The results with $B=10^{18}$ G ignoring the anomalous magnetic moments of nucleons
for IUFSU model are also plotted by dash-dotted line for comparison.
The onset densities of various nonspherical pasta phases are indicated by the circle dots.}
\label{fig:1ea}
\end{figure}
\begin{table}[htb]
\caption{Onset densities of various nonspherical pasta structures and homogeneous matter with different
intensity of magnetic fields $B$ for TM1 and IUFSU models.
The results without the anomalous magnetic moments of nucleons for IUFSU model
are also listed in the last two lines. }
\begin{center}
\begin{tabular}{lcccccc}
\hline\hline
Model & $B$ (G) & \multicolumn{5}{c}{Onset density (fm$^{-3}$)}      \\
\cline{3-7}
      &            &Rod    &Slab   &Tube   &Bubble & Hom.    \\
\hline
TM1   & $0$        & ---   & ---   & ---   & ---   &0.0618  \\
TM1   & $10^{16}$  & ---   & ---   & ---   & ---   &0.0615  \\
TM1   & $10^{17}$  & ---   & ---   & ---   & ---   &0.0610  \\
TM1   & $10^{18}$  &0.0429 & ---   &0.0514 &0.0546 &0.0594  \\
\hline
IUFSU & $0$        &0.0476 &0.0620 &0.0794 &0.0851 &0.0916  \\
IUFSU & $10^{16}$  &0.0476 &0.0632 &0.0770 &0.0851 &0.0916  \\
IUFSU & $10^{17}$  &0.0473 &0.0630 &0.0768 &0.0849 &0.0913  \\
IUFSU & $10^{18}$  &0.0400 &0.0556 &0.0738 &0.0807 &0.0850  \\
\hline
IUFSU ($\kappa_{\rm{p,\,n}}=0$) & $10^{17}$  &0.0473 &0.0631 &0.0768 &0.0851 &0.0916  \\
IUFSU ($\kappa_{\rm{p,\,n}}=0$) & $10^{18}$  &0.0421 &0.0608 &0.0762 &0.0817 &0.0859  \\
\hline\hline
\end{tabular}
\label{tab:3}
\end{center}
\end{table}
\begin{figure}[htb]
\includegraphics[bb=9 40 535 786, width=7 cm,clip]{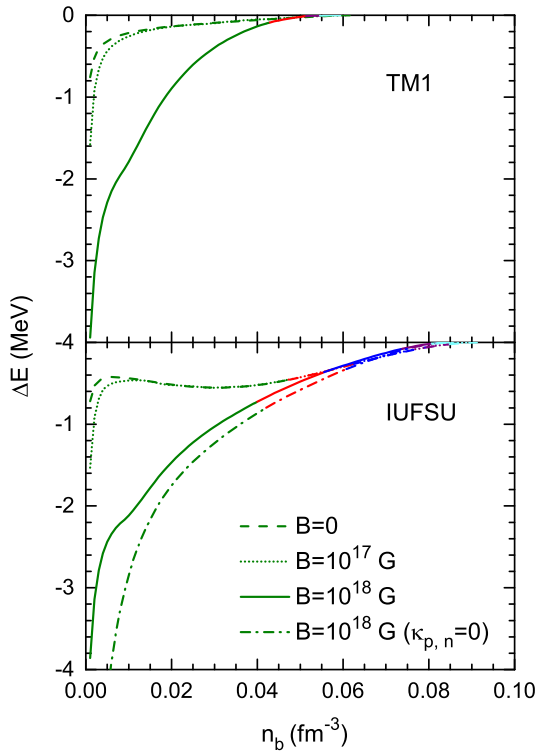}
\caption{(Color online) Same as Fig.~\ref{fig:1ea}, but for
binding energy per nucleon of pasta phases relative to
that of homogeneous matter $\Delta E$. }
\label{fig:2de}
\end{figure}
\begin{figure}[htb]
\includegraphics[bb=5 42 535 786, width=7 cm,clip]{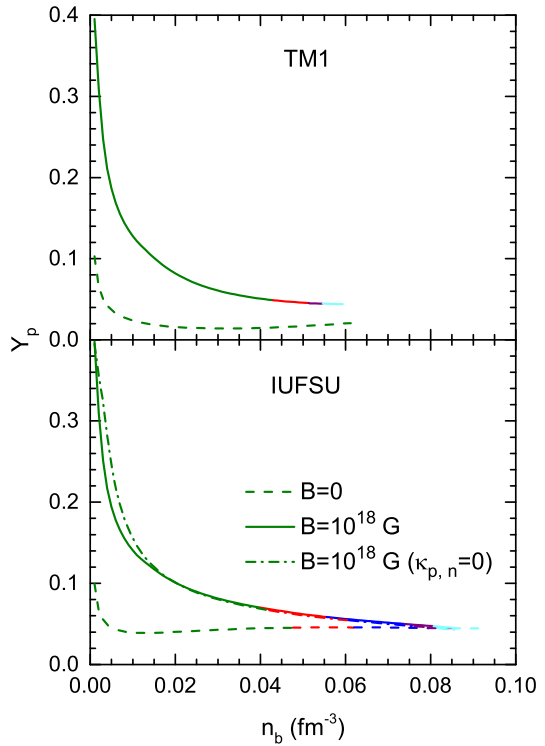}
\caption{(Color online) Proton fractions of pasta phases $Y_p$ as a function of baryon density $n_b$
for TM1 (upper panel) and IUFSU (lower panel) models with magnetic fields $B=10^{18}$ G (solid line)
and $B=0$ (dashed line).
The results with $B=10^{18}$ G ignoring the anomalous magnetic moments of nucleons are also plotted by dash-dotted line for comparison.
Different colors correspond to various pasta structures.}
\label{fig:3yp}
\end{figure}
\begin{figure*}[htb]
\begin{center}
\begin{tabular}{ccc}
\includegraphics[bb=29 19 506 811, width=0.3\linewidth, clip]{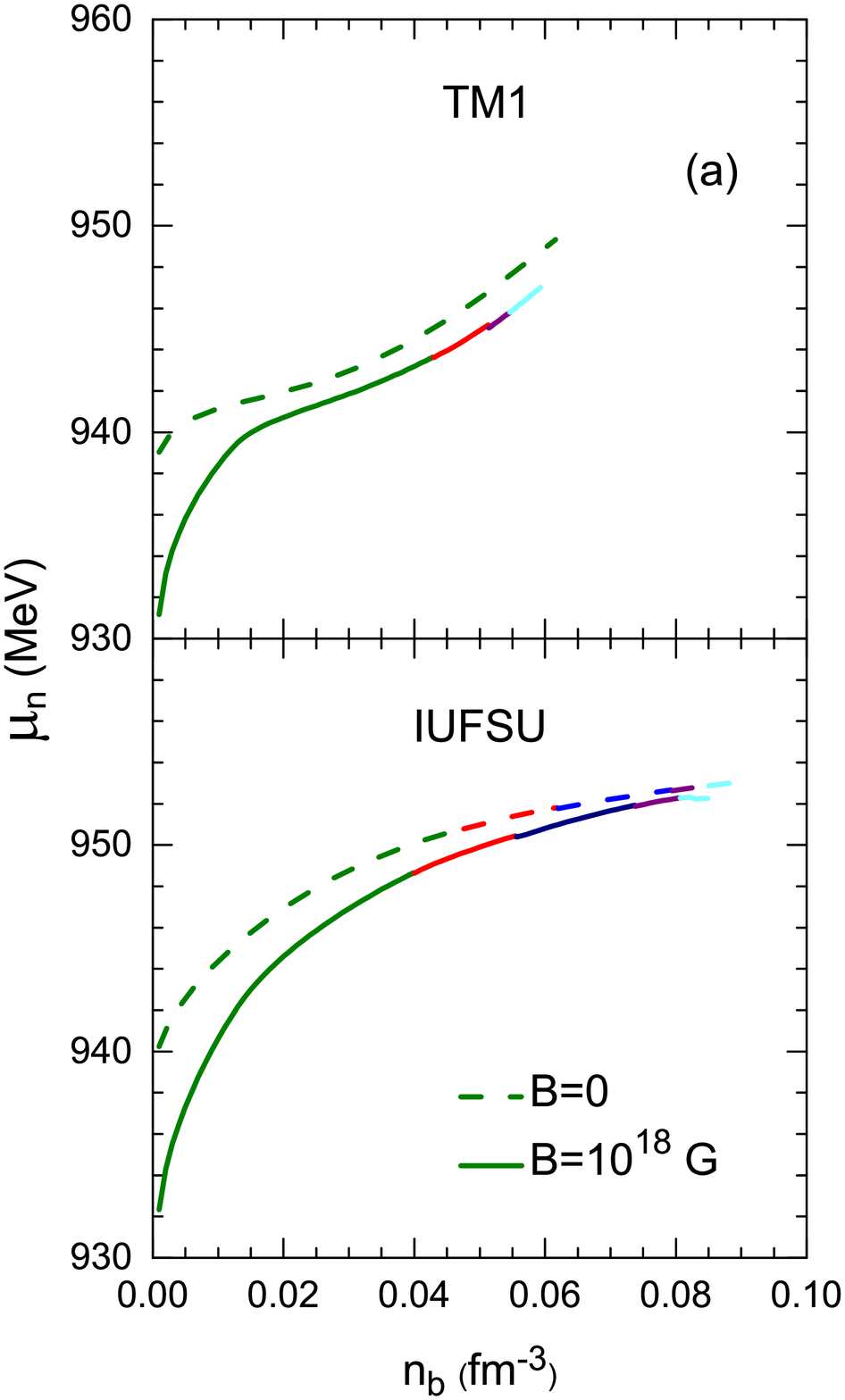}&
\includegraphics[bb=29 19 506 811, width=0.3\linewidth, clip]{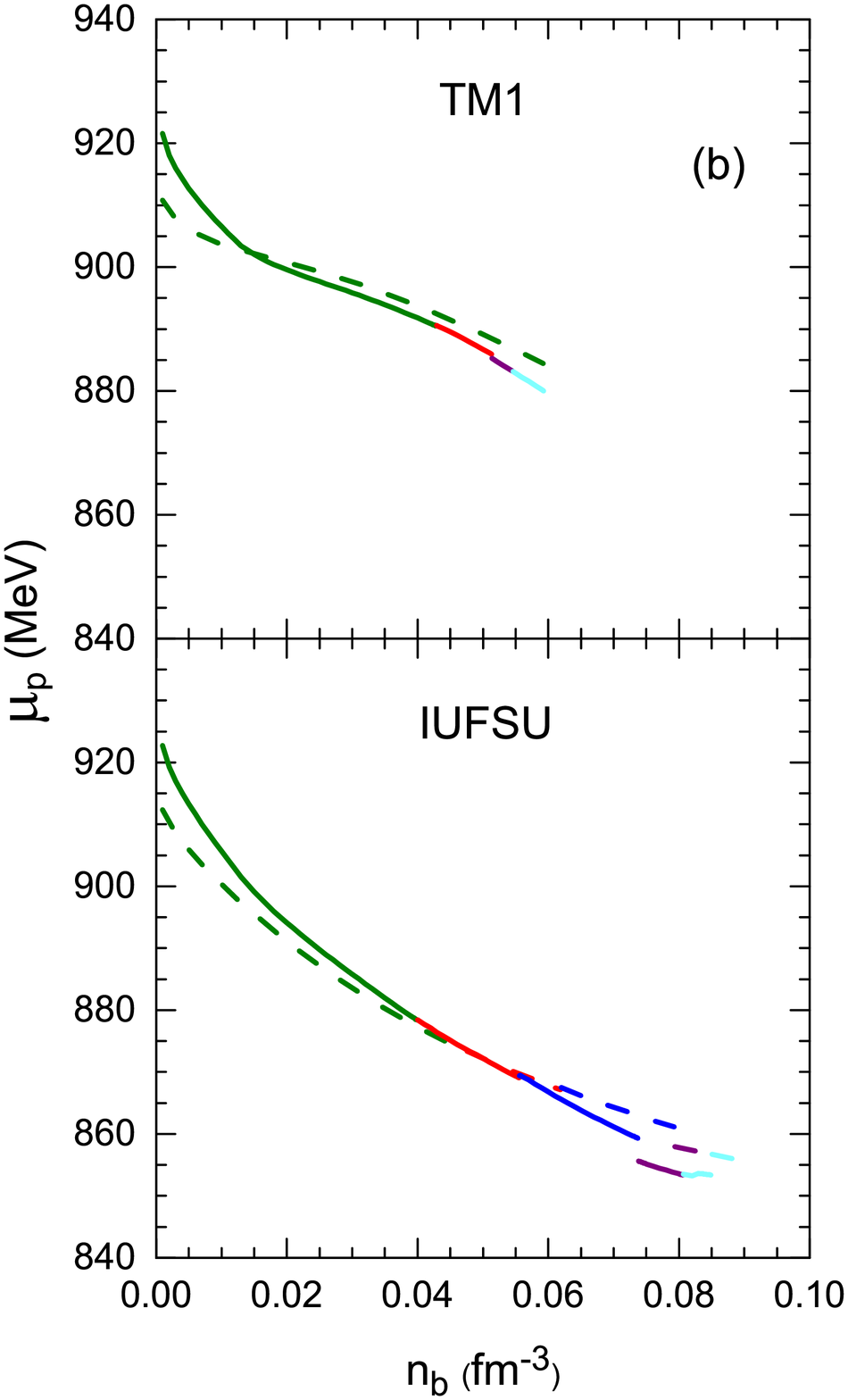}&
\includegraphics[bb=29 19 506 811, width=0.3\linewidth, clip]{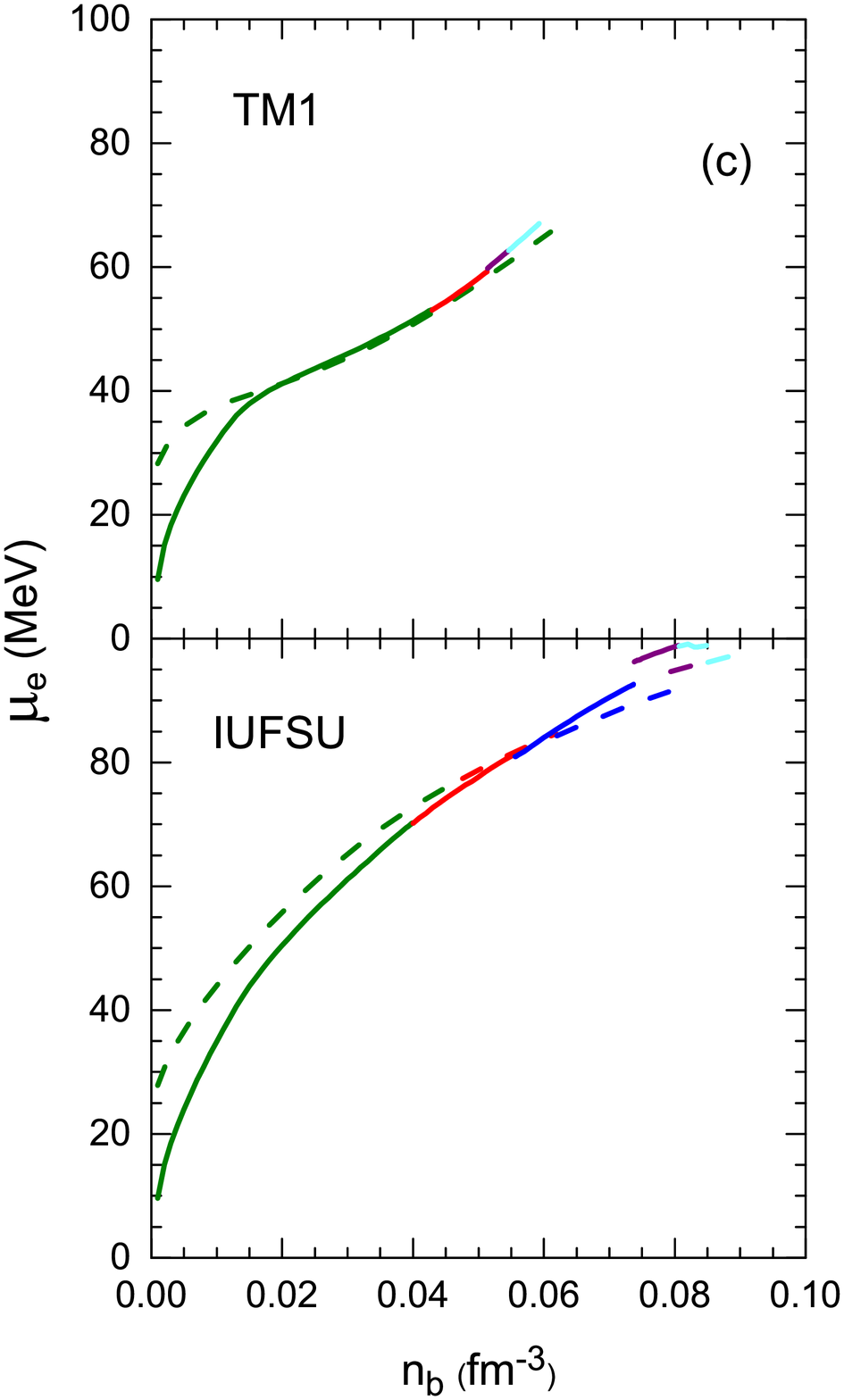}  \\
\end{tabular}
\caption{(Color online) Chemical potentials of neutrons, $\mu_n$ (a), protons, $\mu_p$ (b), and electrons,
$\mu_e$ (c), as functions of baryon density $n_b$ for TM1 (upper panel) and IUFSU (lower panel) models
with magnetic fields $B=10^{18}$ G (solid line) and $B=0$ (dashed line).}
\label{fig:4munpe}
\end{center}
\end{figure*}
\begin{figure}[htb]
\includegraphics[bb=6 39 535 786, width=7 cm,clip]{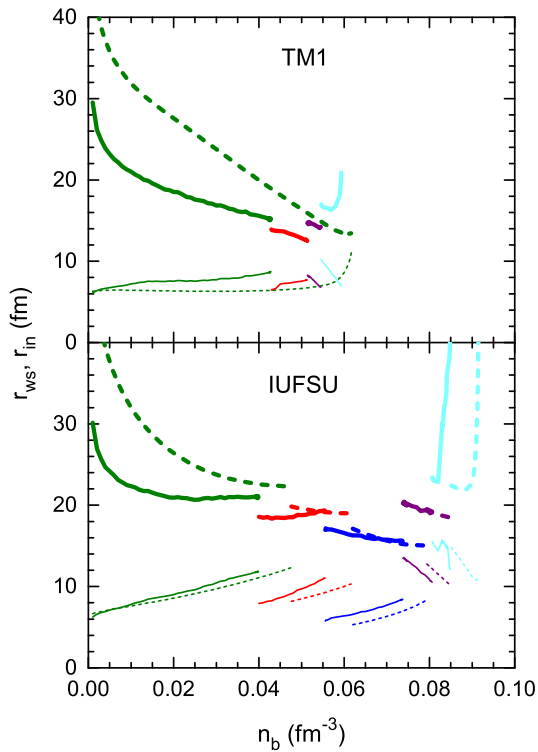}
\caption{(Color online) Radius of WS cell $r_{\rm{ws}}$ (thick line) and nucleus $r_{\rm{in}}$ (thin line)
as a function of baryon density $n_b$ for TM1 (upper panel) and IUFSU (lower panel) models with magnetic fields
$B=10^{18}$ G (solid line) and $B=0$ (dashed line). The jumps in $r_{\rm{ws}}$ and $r_{\rm{in}}$ correspond to
shape transitions in pasta phases.}
\label{fig:5rws}
\end{figure}
\begin{figure*}[htb]
\begin{center}
\begin{tabular}{ccc}
\includegraphics[bb=18 2 564 816, width=7 cm, clip]{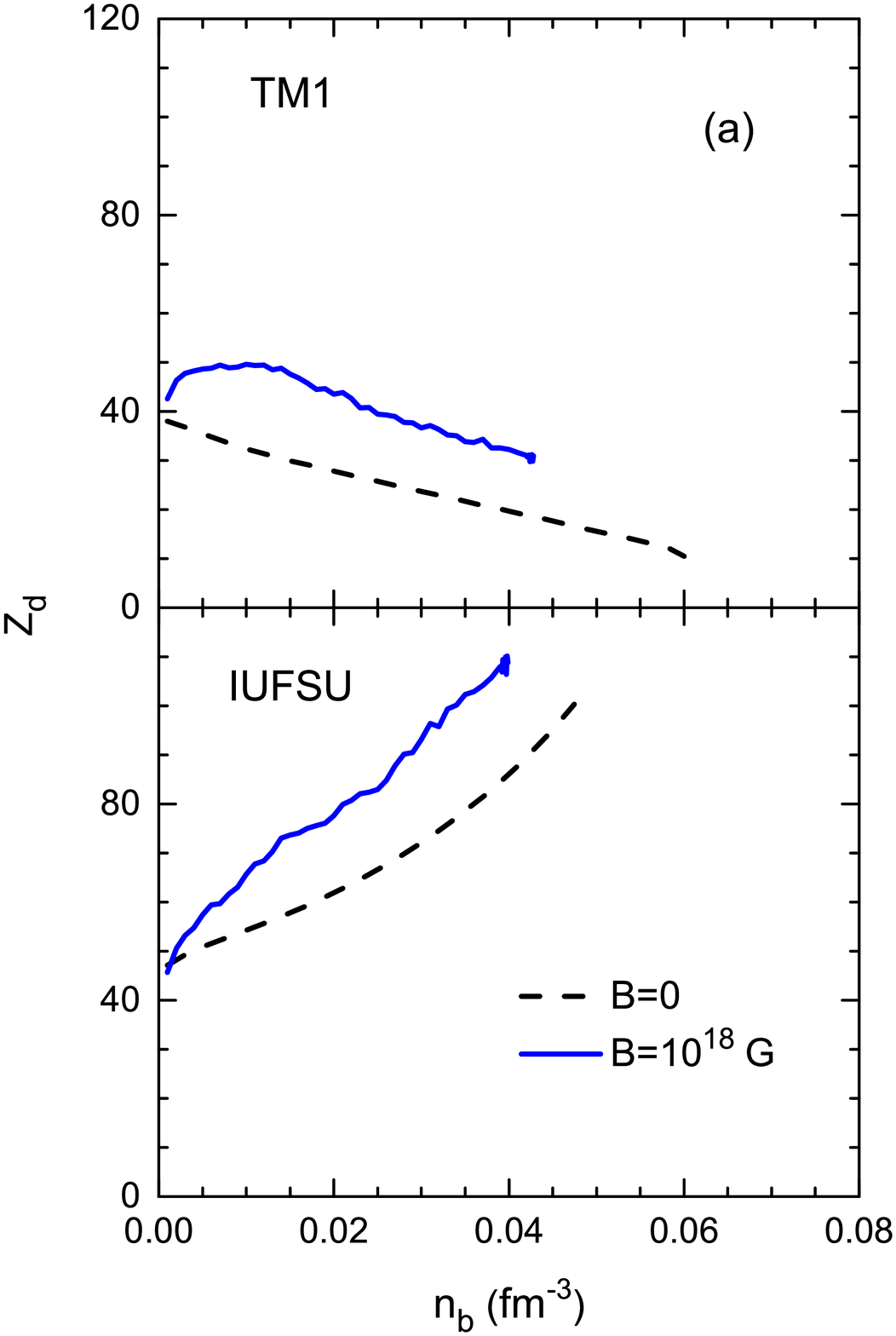}&
\includegraphics[bb=18 2 564 816, width=7 cm, clip]{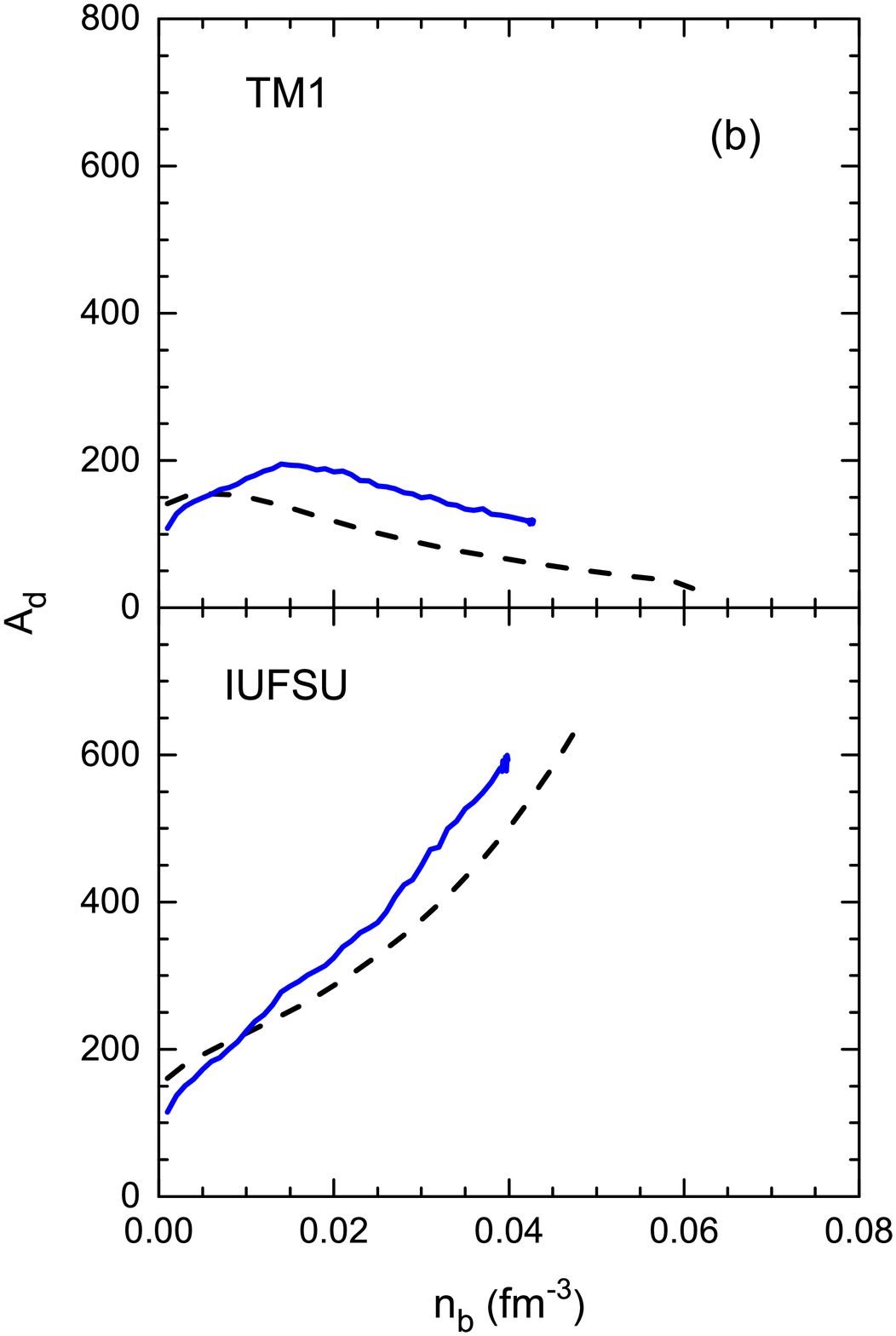}  \\
\end{tabular}
\caption{(Color online) Properties of spherical nuclei in droplet phase, such as the
charge number $Z_d$ (a) and the nucleon number $A_d$ (b),
as a function of baryon density $n_b$ for TM1 (upper panel) and IUFSU (lower panel) models with magnetic fields $B=0$ G (dashed line) and $B=10^{18}$ G (solid line).}
\label{fig:6zad}
\end{center}
\end{figure*}
\begin{figure*}[htb]
\begin{center}
\begin{tabular}{ccc}
\includegraphics[bb=24 33 429 767, width=0.3\linewidth, clip]{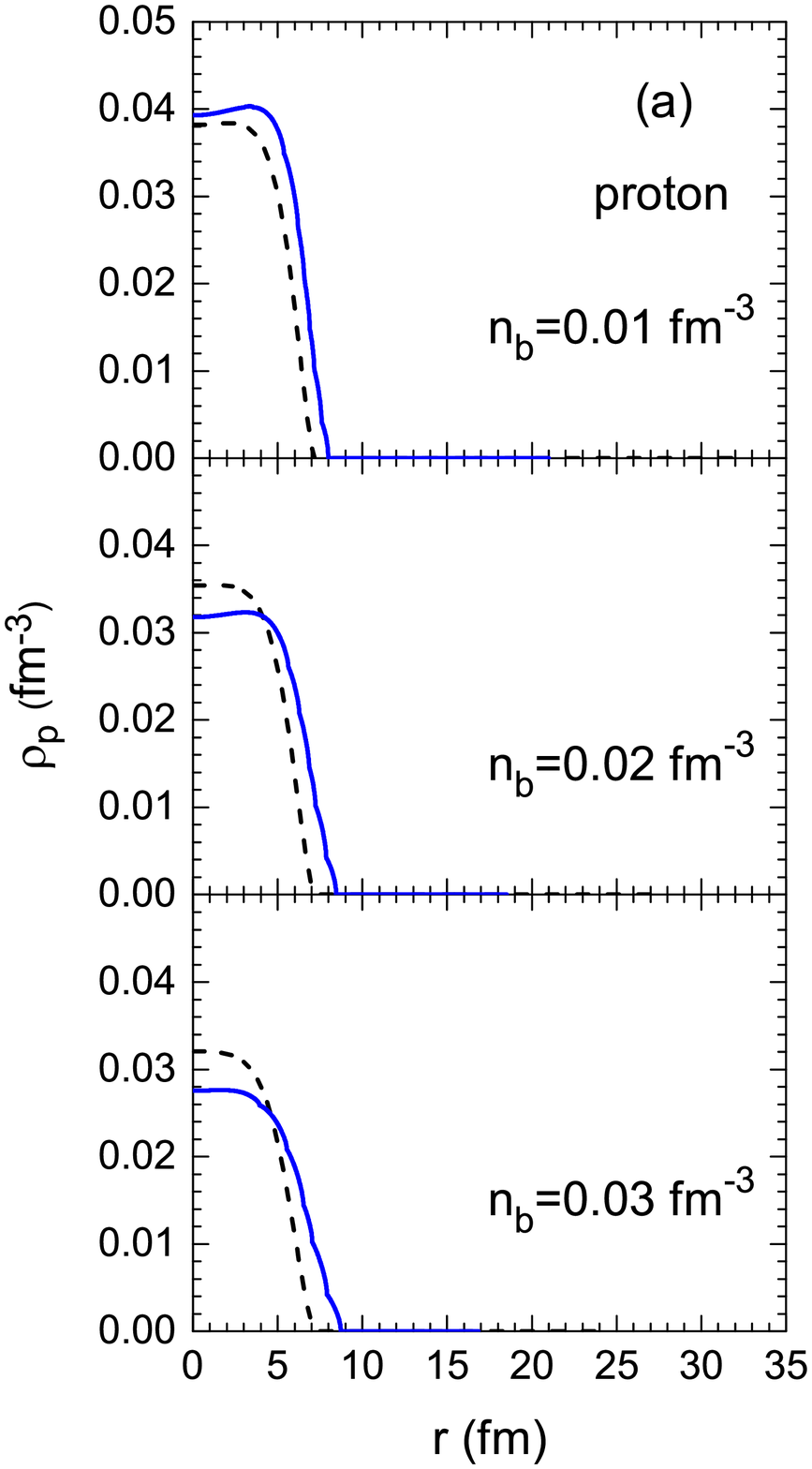}&
\includegraphics[bb=24 33 429 767, width=0.3\linewidth, clip]{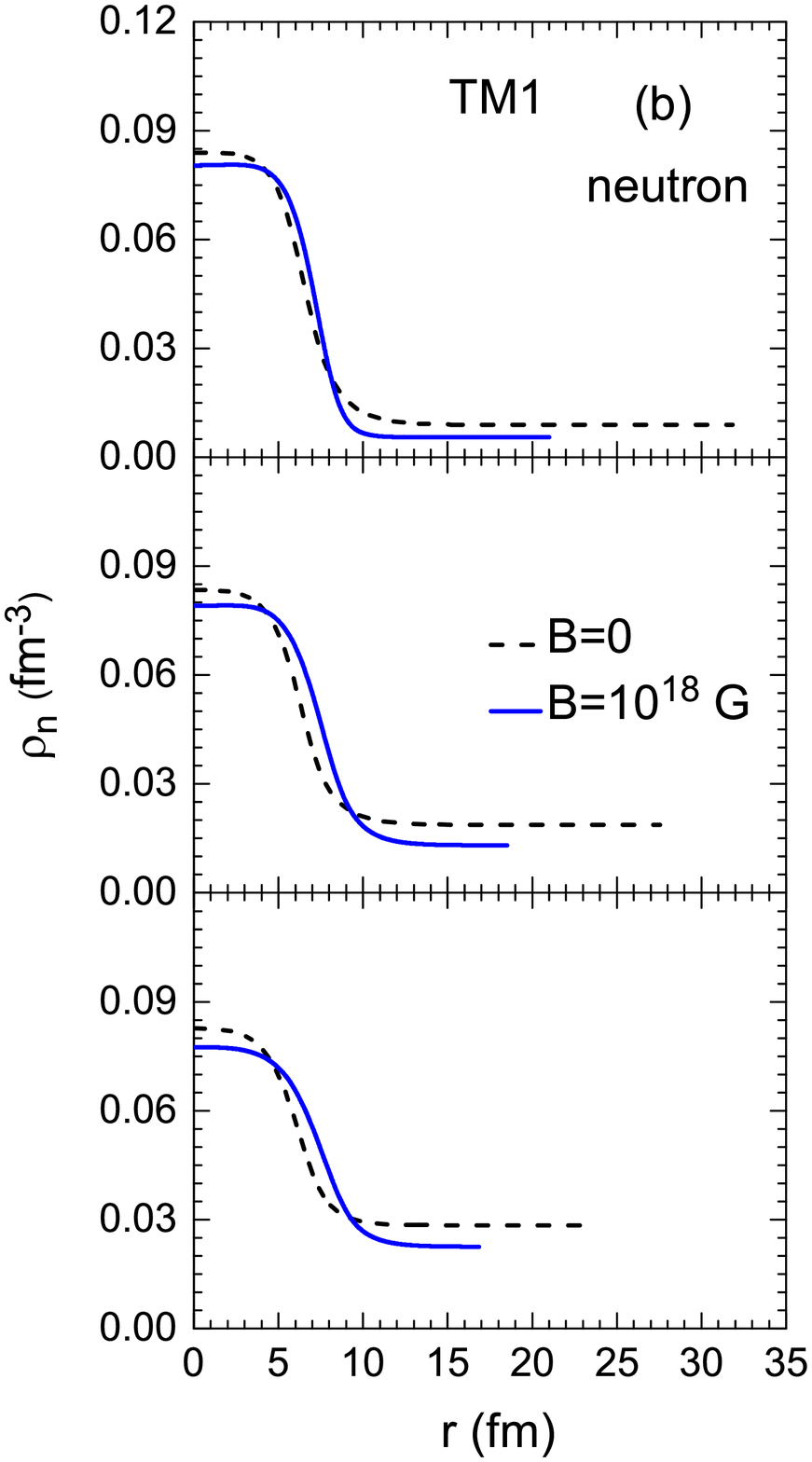}&
\includegraphics[bb=24 33 429 767, width=0.3\linewidth, clip]{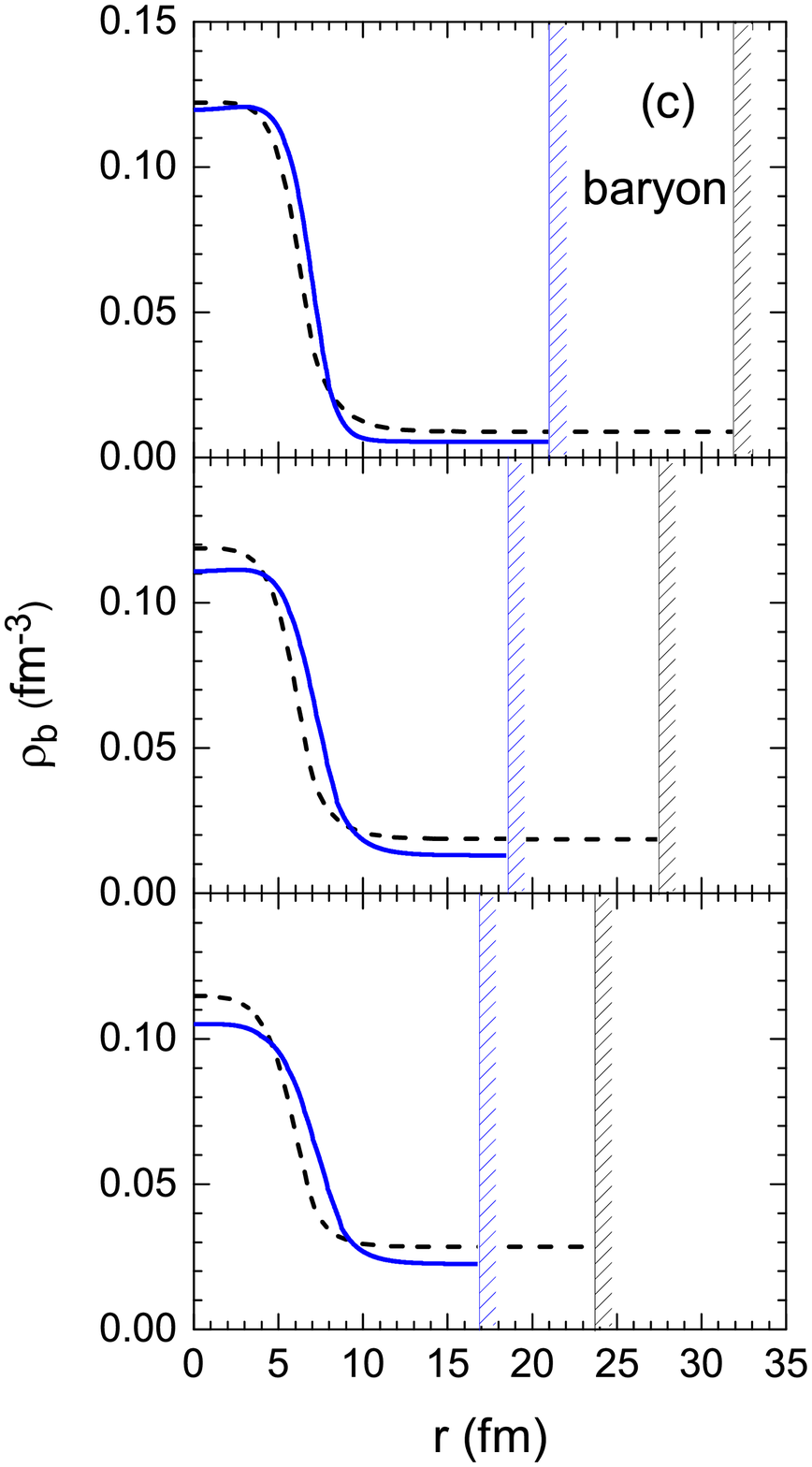}  \\
\end{tabular}
\caption{(Color online) Density distributions of protons, $\rho_p$ (a), neutrons, $\rho_n$ (b), and baryons,
$\rho_b$ (c), in the WS cell at $n_b=0.01,\ 0.02,\ 0.03$ fm$^{-3}$ (top to bottom) obtained in the
TF approximation for TM1 model with magnetic fields $B=0$ (dashed line) and $B=10^{18}$ G (solid line). The cell boundary is indicated by the hatching.}
\label{fig:7tmpnb}
\end{center}
\end{figure*}
\begin{figure*}[hbt]
\begin{center}
\begin{tabular}{ccc}
\includegraphics[bb=24 33 429 767, width=0.3\linewidth, clip]{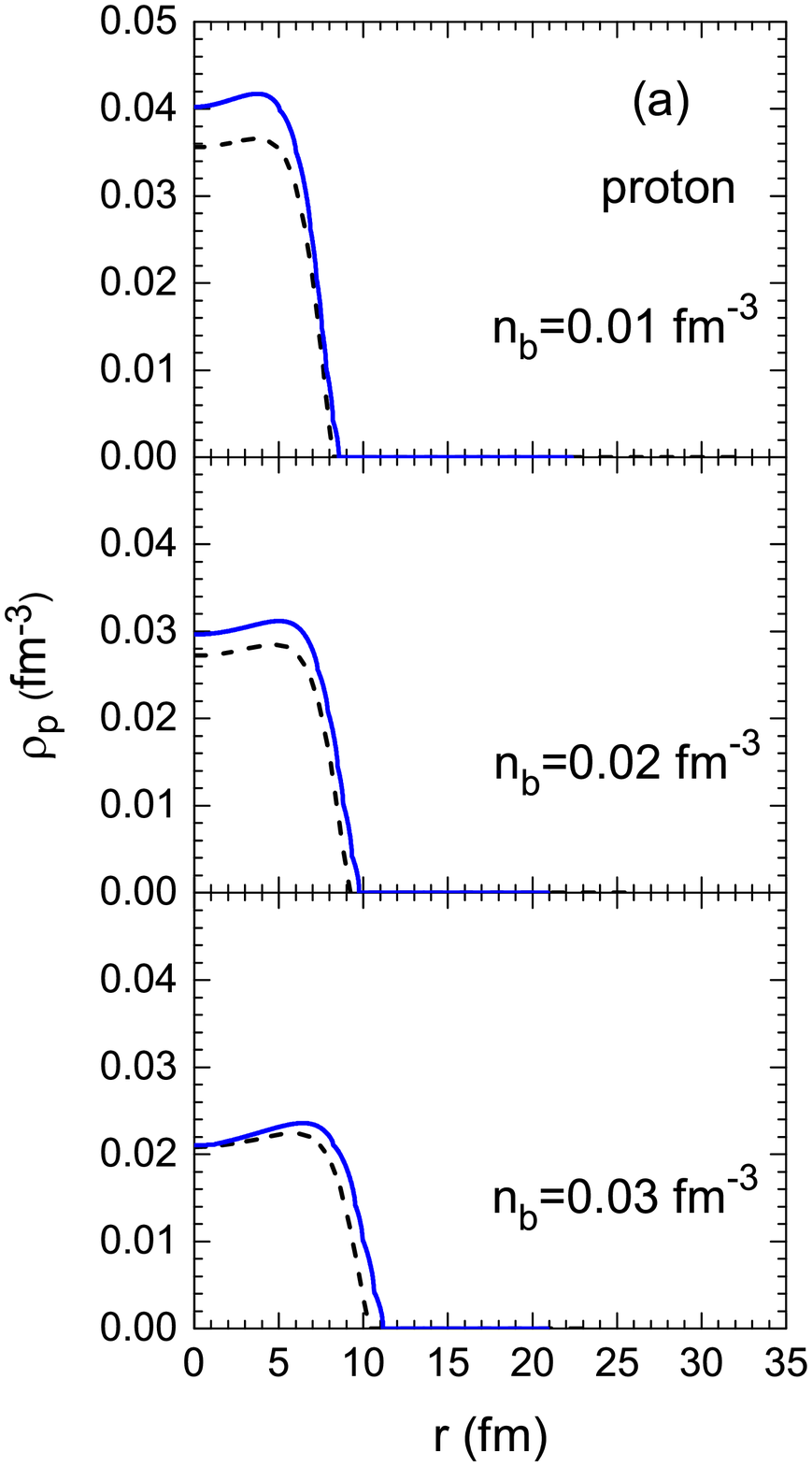}&
\includegraphics[bb=24 33 429 767, width=0.3\linewidth, clip]{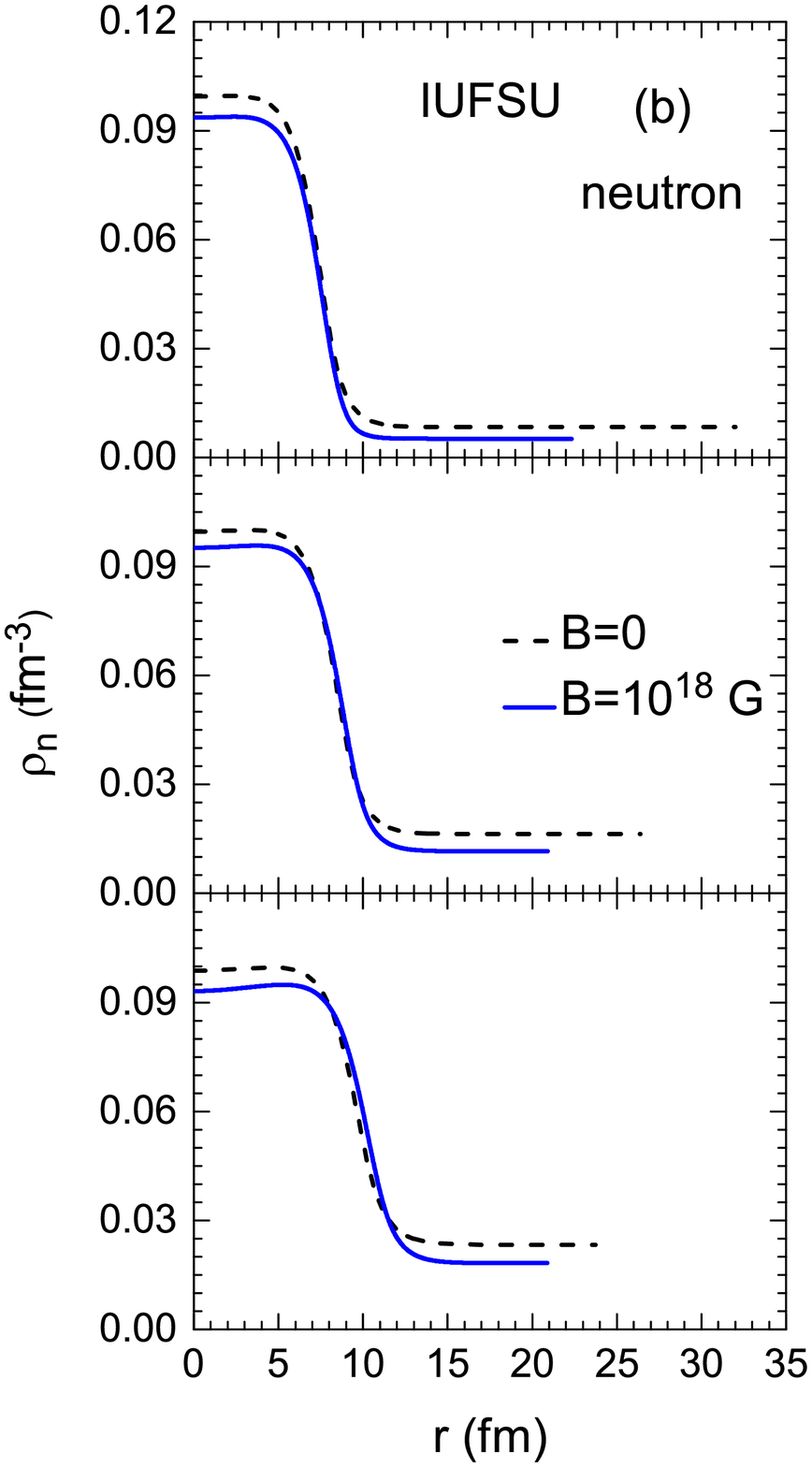}&
\includegraphics[bb=24 33 429 767, width=0.3\linewidth, clip]{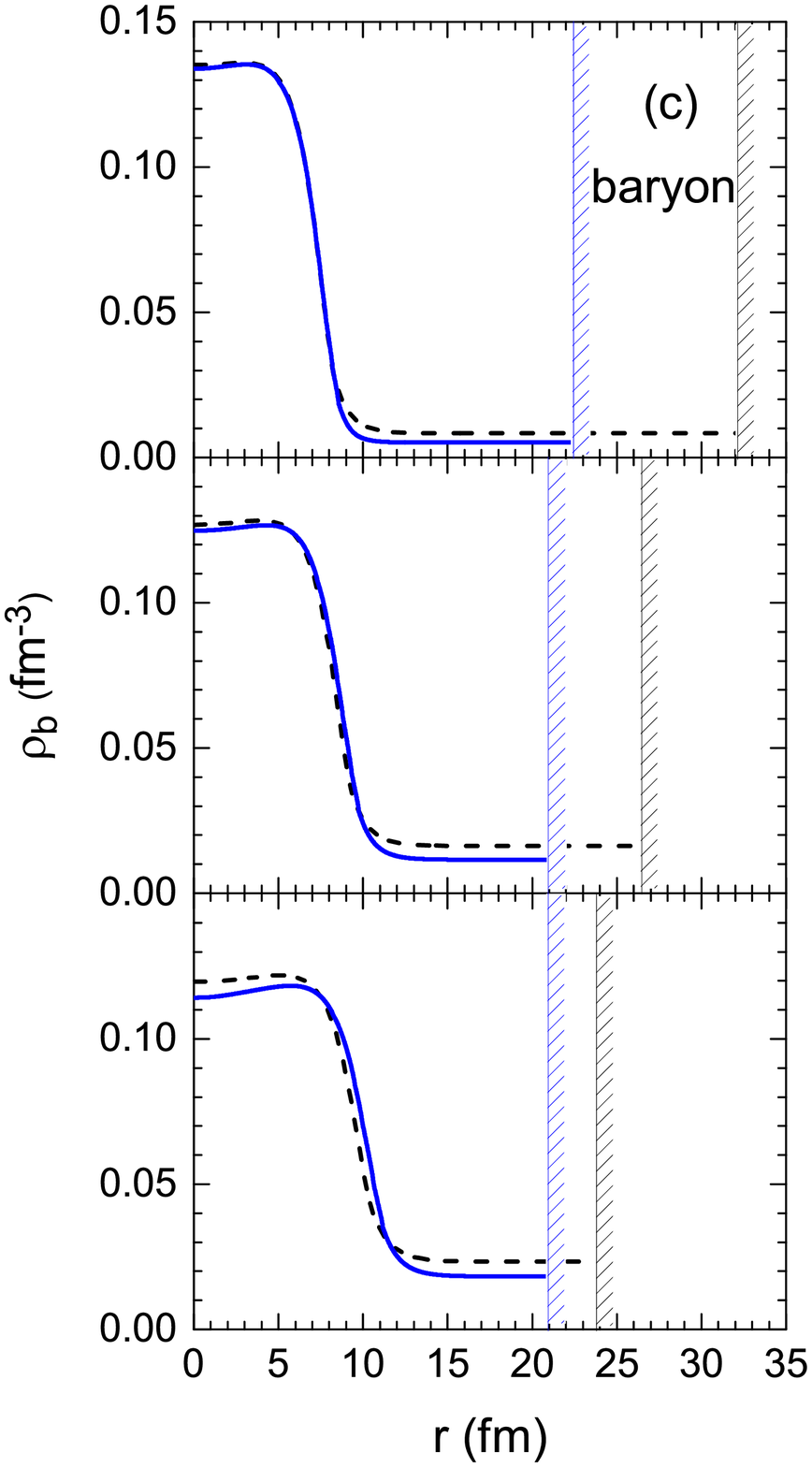}  \\
\end{tabular}
\caption{(Color online) Same as Fig.~\ref{fig:7tmpnb}, but for IUFSU model.}
\label{fig:8iupnb}
\end{center}
\end{figure*}
\begin{figure*}[htb]
\begin{center}
\begin{tabular}{ccc}
\includegraphics[bb=33 8 381 809, width=0.3\linewidth, clip]{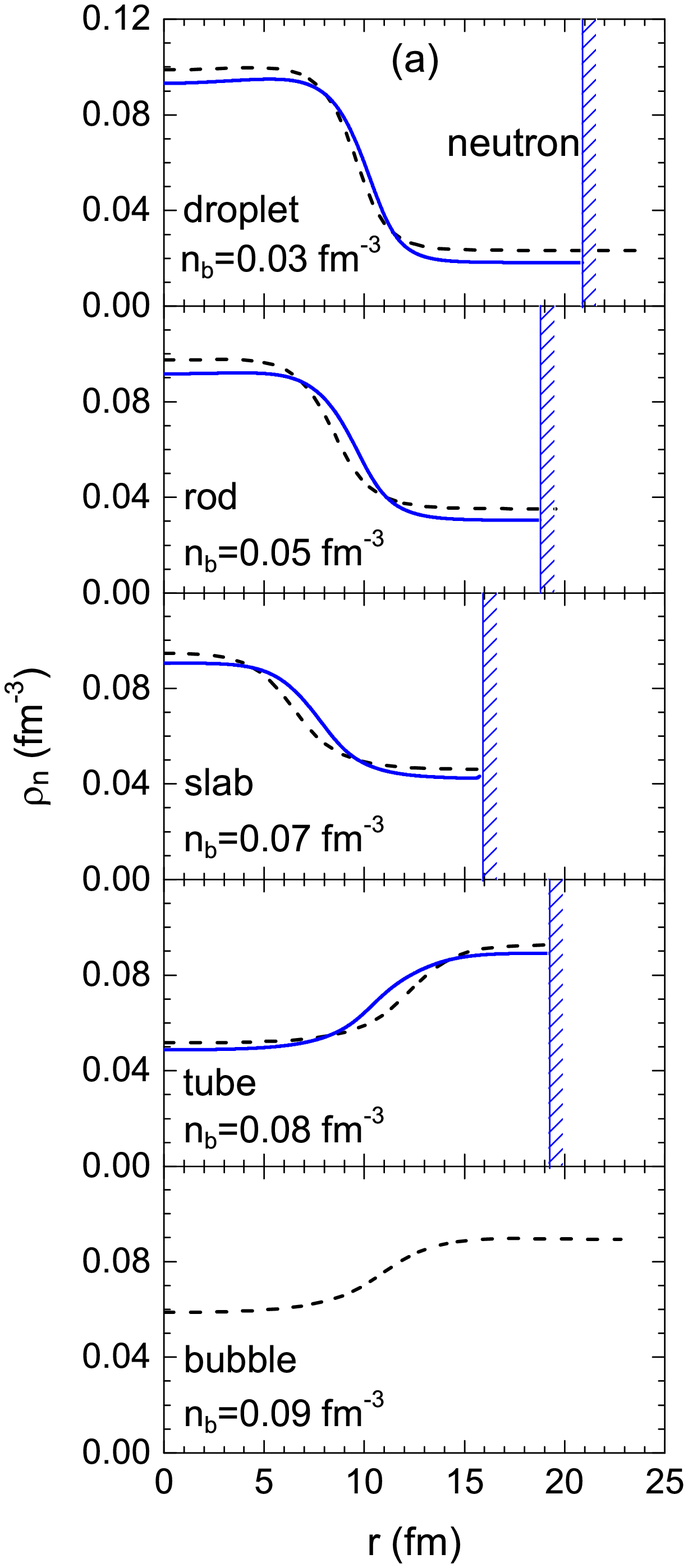}&
\includegraphics[bb=33 8 381 809, width=0.3\linewidth, clip]{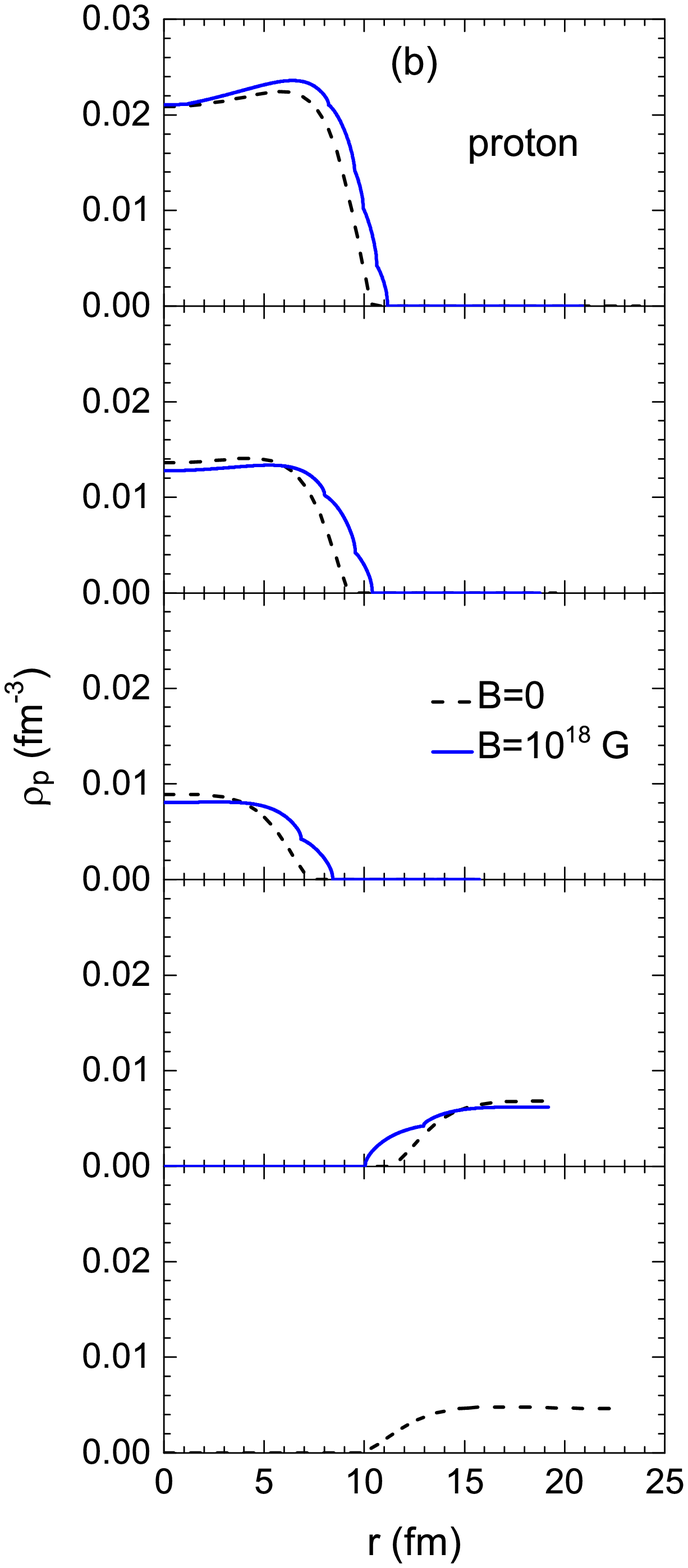}&
\includegraphics[bb=33 8 381 809, width=0.3\linewidth, clip]{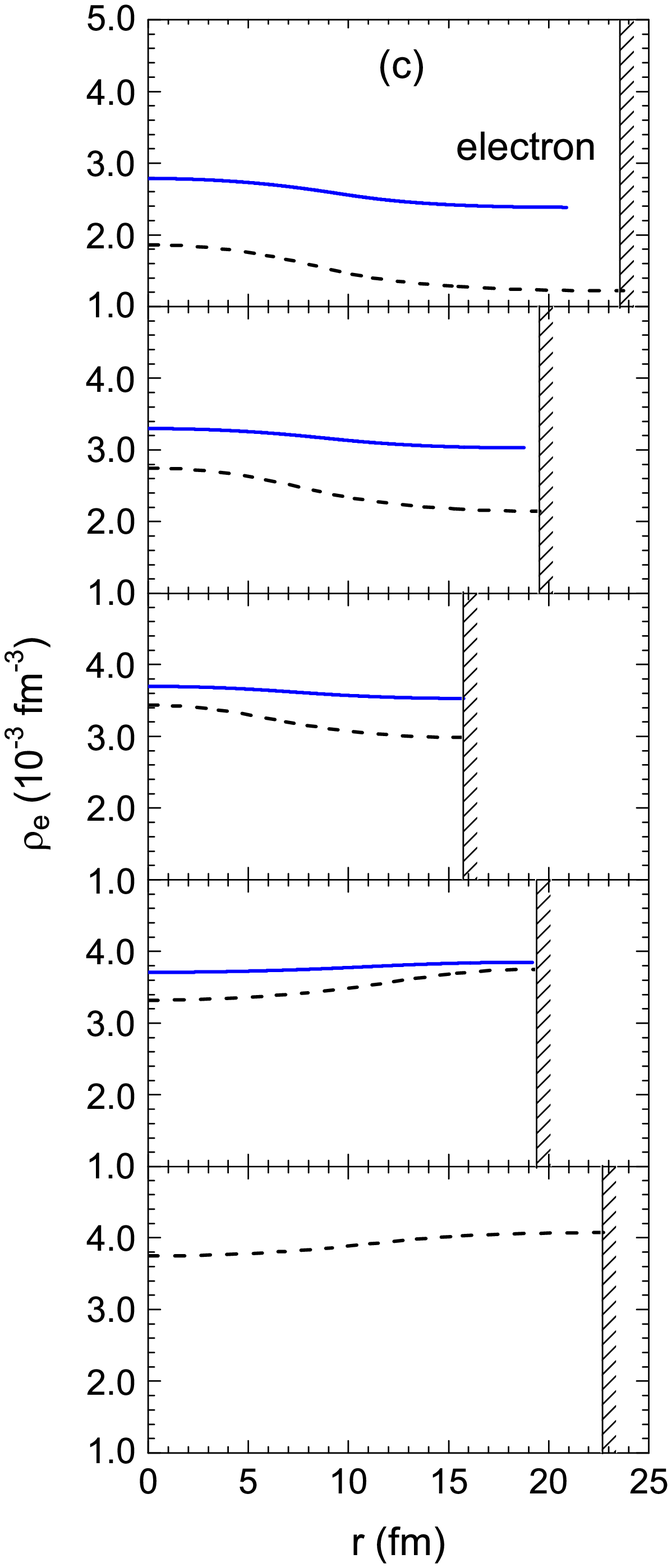}  \\
\end{tabular}
\caption{(Color online) Density distributions of neutrons, $\rho_n$ (a), protons, $\rho_p$ (b), and electrons,
$\rho_e$ (c), in the WS cell at $n_b=0.03,\ 0.05,\ 0.07,\ 0.08,\, 0.09$ fm$^{-3}$ (top to bottom) obtained in the
TF approximation for IUFSU model with magnetic fields $B=0$ (dashed line) and $B=10^{18}$ G (solid line). The cell boundary is indicated by the hatching.}
\label{fig:9npe}
\end{center}
\end{figure*}
\begin{figure*}[htb]
\begin{center}
\begin{tabular}{cc}
\includegraphics[bb=5 121 587 642, width=0.4\linewidth, clip]{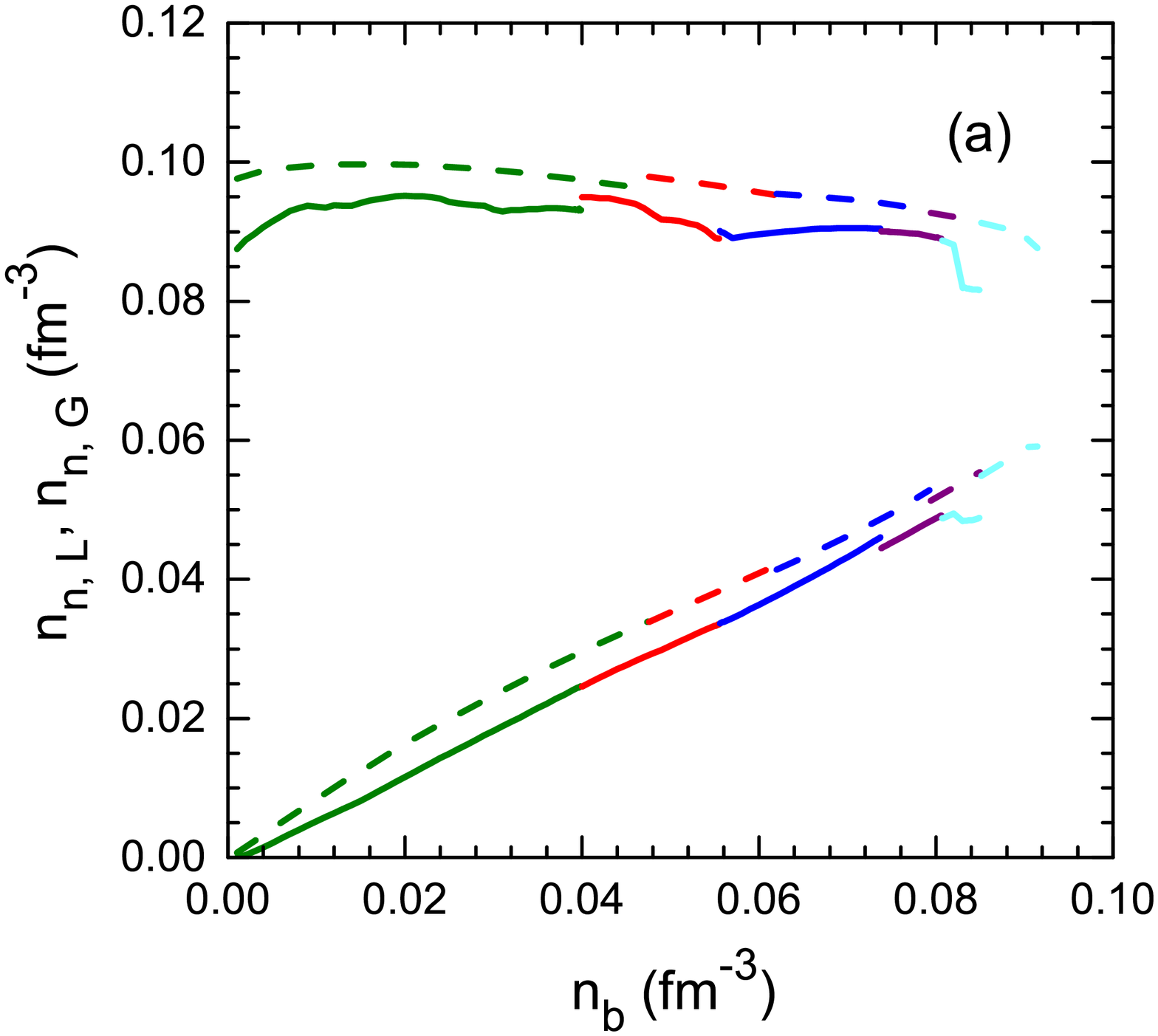}&
\includegraphics[bb=5 121 587 642, width=0.4\linewidth, clip]{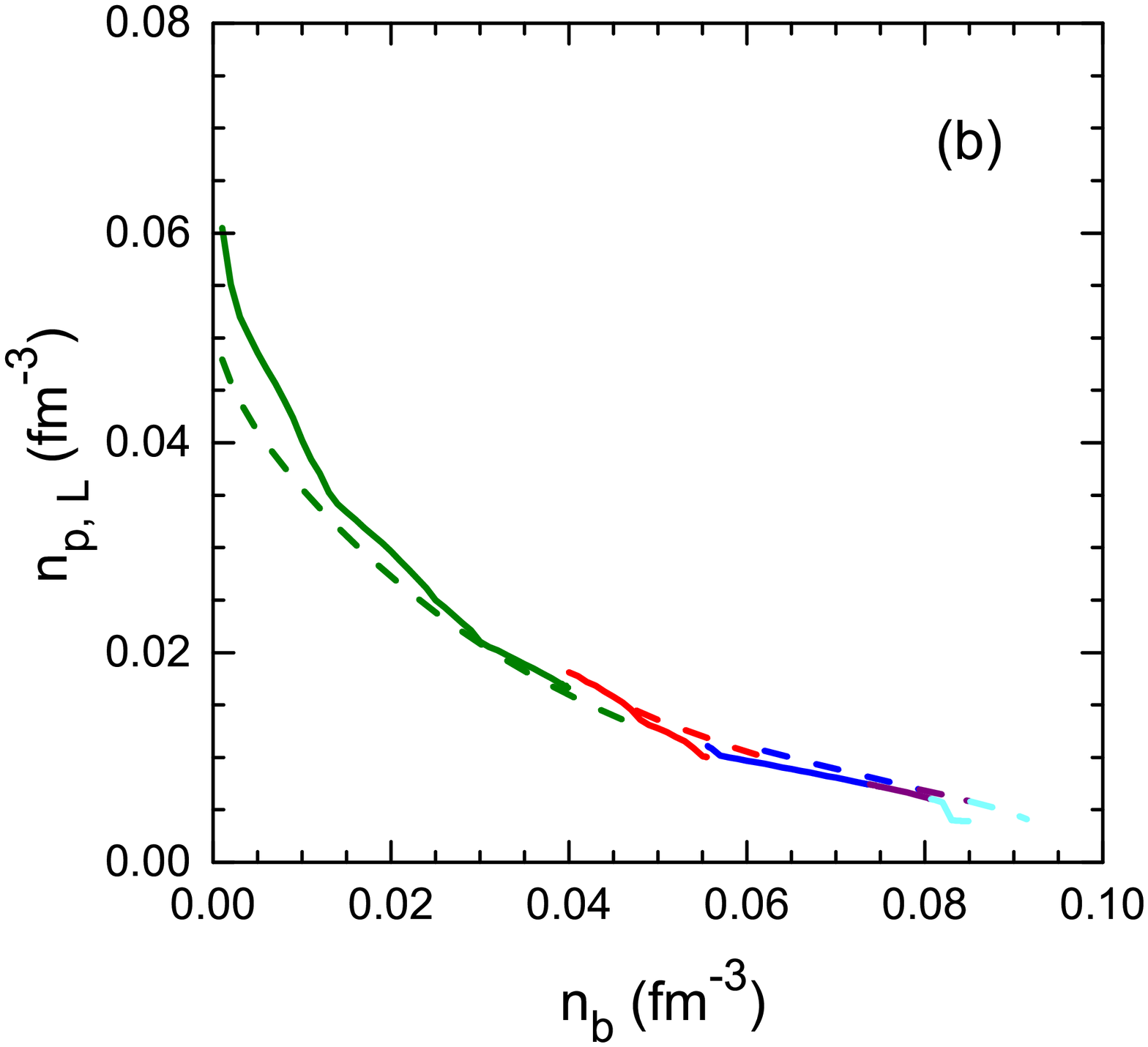}  \\
\end{tabular}
\caption{(Color online) Neutron density of liquid phase $n_{n,\,L}$ and gas phase $n_{n,\,G}$ in WS cell (a)
and proton density of liquid phase $n_{p,\,L}$ in WS cell (b), as a function of baryon density $n_b$
for IUFSU model with magnetic fields $B=0$ (dashed line) and $B=10^{18}$ G (solid line).
The kinks correspond to shape transitions of different pasta phases.}
\label{fig:10nnnp}
\end{center}
\end{figure*}

We employ the WS approximation to describe the inner crust structure of neutron star,
assuming that only one nucleus is included in a WS cell,
where the nucleus coexists with neutron and surrounded by electron gases.
The $\beta$ equilibrium and charge neutrality conditions are satisfied in a WS cell,
\begin{eqnarray}
\mu _{n} &=&\mu _{p}+\mu _{e},
\label{eq:beta} \\
N_{e} &=&N_{p},
\label{eq:charge}
\end{eqnarray}
where the chemical potentials of nucleons and electron are written as
\begin{eqnarray}
\mu _{n} &=&{E_{F}^{n}}+g_{\omega}\omega -\frac{g_{\rho }}{2}\rho, \\
\label{eq:mun}
\mu _{p} &=&{E_{F}^{p}}+g_{\omega}\omega +\frac{g_{\rho }}{2}\rho +e A, \\
\label{eq:mup}
\mu _{e} &=&{E_{F}^{e}} - e A,
\label{eq:mue}
\end{eqnarray}%
and the numbers of electrons and protons  inside the WS cell are given by
\begin{eqnarray}
N_{e}=\int_{\rm{cell}}n_{e}(r)d^{3}r,  \\
\label{eq:necell}
N_{p}=\int_{\rm{cell}}n_{p}(r)d^{3}r.
\label{eq:npcell}
\end{eqnarray}
At a given average baryon density $n_b$ as well as radius of WS cell $r_{\rm{ws}}$,
we adopt the TF approximation to calculate the distributions of nucleons and electrons.
In practice, we start with an initial guess for meson fields
$\sigma(r)$, $\omega(r)$, $\rho(r)$, and electromagnetic field $A(r)$,
and then determine the chemical potentials, $\mu_n$, $\mu_p$, and $\mu_e$
under the constraints of Eqs.~(\ref{eq:beta}) and (\ref{eq:charge}) and
baryon number conservation,
\begin{eqnarray}
n_bV_{\rm{cell}}=\int_{\rm{cell}}\left[n_{p}(r)+n_{n}(r)\right]d^{3}r.
\end{eqnarray}
Once the chemical potentials are determined, it is easy to calculate various densities and
new mean fields by solving Eqs.~(\ref{eq:eqms})--(\ref{eq:eqma}).
This procedure should be iterated until convergence is achieved.
Furthermore, we calculate the total energy of WS cell
\begin{equation}
E_{\rm{cell}}=\int_{\rm{cell}}{\varepsilon }_{\rm{rmf}}(r)d^{3}r,
\label{eq:TFe}
\end{equation}%
and binding energy per nucleon
\begin{equation}
E/N=\frac{E_{\rm{cell}}}{n_bV_{\rm{cell}}}-M.
\end{equation}
We consider five nuclear pasta structures in this work.
The volume of WS cell for different pasta shapes is given by
\begin{equation}
V_{\rm{cell}}=\left\{
\begin{array}{ll}
\frac{4}{3}{\pi}r_{\text{ws}}^{3}, & \text{for droplet and bubble},  \\
l{\pi}r_{\text{ws}}^{2},           & \text{for rod and tube},        \\
2{l^2}r_{\text{ws}},               & \text{for slab},
\end{array}
\right.
\label{eq:vcell}
\end{equation}
where $l$ is the length for rod and tube and $l$ is the width for slab.
We notice that the value of $l$ does not affect the binding energy per nucleon $E/N$ and
is somewhat arbitrary.

At a given average baryon density $n_b$, we minimize the binding energy per nucleon $E/N$ with
respect to the cell size $r_{\rm{ws}}$ for all five pasta configurations
and then we compare $E/N$ between different configurations in order to determine the most
stable shape that has the lowest $E/N$.
Besides, the binding energy per nucleon of homogeneous matter at the same $n_b$ is also calculated
and compared to determine the crust-core transition where $E/N$ of homogeneous matter becomes lower
than that of stable pasta phase.
In the TF approximation, there is no distinct boundary between the dense nuclear phase and the dilute gas phase, so we prefer to adopt the definition in Ref.~\cite{Bao15},
\begin{equation}
r_{\rm{in}}=\left\{
\begin{array}{ll}
r_{\rm{ws}} \left(\frac{\langle n_{p}\rangle^{2}}{\langle n_{p}^{2}\rangle}\right)^{1/D},
& \text{for droplet, rod, and slab}, \\
r_{\rm{ws}}\left(1-\frac{\langle n_{p}\rangle ^{2}}{\langle n_{p}^{2}\rangle}\right)^{1/D},
& \text{for tube and bubble},%
\end{array}%
\right.
\end{equation}
to measure the size of inner part in the WS cell,
where the average values in brackets $\left<\cdots\right>$ are calculated over the cell volume $V_{\rm{cell}}$
and the dimension of WS cell $D=1,2,3$ for slab, rod (tube), and droplet (bubble), respectively.


\section{Results and discussion}
\label{sec:3}

In this section, we show the numerical results obtained by using self-consistent TF approximation
and discuss the effects of strong magnetic fields on the properties of neutron star crust.
The results obtained with different intensity of magnetic fields
in TM1 model are compared with that in IUFSU model.
The parameter sets and saturation properties of these two RMF models are given in
Tables~\ref{tab:1} and~\ref{tab:2}, respectively.
In Fig.~\ref{fig:1ea},
we plot the binding energy per nucleon $E/N$ of pasta phases as a
function of average baryon density $n_b$ for TM1 (upper panel) and IUFSU (lower panel) models
with and without strong magnetic fields.
We can see that the binding energy $E/N$ with $B=10^{17}$ G is slightly smaller than the one with $B=0$,
while both of them are obviously larger than that with $B=10^{18}$ G.
This behavior is consistent with the results in Ref.~\cite{Lima13}.
It is because the existence of large degeneracy of the Landau levels in strong magnetic fields
can soften the equation of state.
We also notice that only the droplet configuration exists
as $B\leq10^{17}$ G before the crust-core transition in the case of the TM1 model.
However, all pasta phases arise whether the magnetic fields are considered in the case of IUFSU model.
In order to check the effects of anomalous magnetic moments of nucleons on pasta phase,
we also calculate the pasta structures for different strength of magnetic field with $\kappa_{\rm{p,\,n}}=0$ in the IUFSU model.
It is found that the anomalous magnetic moments of nucleons have very little impact on pasta structure as $B\leq10^{17}$ G,
while it should not be neglected as $B\cong10^{18}$ G, which is also plotted in Fig.~\ref{fig:1ea} for comparison.
One can see that $E/N$ with $\kappa_{\rm{p,\,n}}=0$ is obviously larger than the result
with the inclusion of anomalous magnetic moments.
Besides, the onset densities of pasta phases are also changed.

The transition densities of various pasta phases and crust-core transition density with different
intensity of magnetic fields are listed in detail in Table~\ref{tab:3}.
It is found that the results with $B=10^{17}$ G for IUFSU model do not change much
whether the anomalous magnetic moments of nucleons are considered.
However, for $B=10^{18}$ G, considerable differences are observed in the onset densities of nonspherical pasta phases and the transition density to homogeneous matter.
For both TM1 and IUFSU models, one can see that the results with $B=10^{16}$ G are quite similar to those with $B=0$,
so the effects of magnetic fields on pasta structures can be neglected
when the strength of magnetic fields $B$ is not larger than $\cong10^{16}$ G.
So, we will not discuss the results with $B=10^{16}$ G in the following contents.
Comparing the results of TM1 and IUFSU models, the pasta structures are significantly different for various values of $B$.
In the TM1 model, the nonspherical structures such as rod, tube, and bubble appear only
in the case of $B=10^{18}$ G; however, the slab structure is absent.
In the IUFSU model, all five kinds of pasta structures occur with and without strong magnetic fields.
The differences between these two models should be due to their different symmetry energy and its density dependence.
It has been found that a smaller symmetry energy slope could result in more complex pasta phases~\cite{Bao15}.
On the other hand, as $B$ increases, the onset density of homogeneous matter,
namely the crust-core transition density, decreases both in TM1 and IUFSU models.
The transition densities between different pasta phases also decrease with increasing $B$ as observed in the IUFSU model.
We also notice that the transition density at the bubble-homogeneous matter is nonmonotonic with increasing $B$ in Ref.~\cite{Lima13} using NL3 parametrization to perform the calculation,
where the proton fraction is fixed as $Y_p=0.3$.
This value is much larger than the results of $\beta$ equilibrium in this work.

The behaviors in Table~\ref{tab:3} can be understood from Fig.~\ref{fig:2de},
where we plot the differences between the binding energy per nucleon of pasta phase and
that of homogeneous matter $\Delta E$ as a function of baryon density $n_b$ with $B=0,\, 10^{17},\, 10^{18}$ G.
We can see that a larger $B$ results in a smaller $\Delta E$
at lower baryon densities and the results with $B=10^{18}$ G are much lower than
those with $B=0,\, 10^{17}$ G.
However, as $n_b$ increases, $\Delta E$ with $B=10^{18}$ G raises rapidly
and then exceeds the results with $B=0,\, 10^{17}$ G.
As a result, $\Delta E$ with larger $B$ reaches ``$\Delta E=0$'' earlier, which leads to a smaller
crust-core transition density.

In Fig.~\ref{fig:3yp}, we plot the proton fraction of pasta phase $Y_p$ with $B=0,\, 10^{18}$ G
for TM1 and IUFSU models.
The results with $B=10^{16},\, 10^{17}$ G will not be shown, considering
no obvious differences from the results with $B=0$.
The results with $B=10^{18}$ G neglecting the anomalous magnetic moments of nucleons for IUFSU model are also plotted.
One can see that the proton fraction for $\kappa_{\rm{p,\,n}}=0$ is slightly  larger than that, including anomalous magnetic moments.
It can be understood from Eqs.~(\ref{eq:epf}) and~(\ref{eq:enf}).
For $\kappa_{\rm{p,\,n}}=0$, the proton Fermi energy $E_F^p$ decreases while the neutron Fermi energy $E_F^n$ increases,
which leads to more proton energy levels occupied.
We can see in Fig.~\ref{fig:3yp} that the proton fraction $Y_p$ with $B=10^{18}$ G
is much larger than the results with $B=0$,
especially at lower densities.
It can be understood from Eq.~(\ref{eq:np}).
We notice that only the zeroth Landau level is occupied,
and $eB$ is much larger than $k^{p2}_{F,\nu,s}$ at lower densities,
when the magnetic field $B=10^{18}$ G is included.
As a result, the proton fraction $Y_p$ with $B=10^{18}$ G is larger than that with $B=0$.
As $n_b$ increases, $k_{F,\nu,s}^{p}$ increases rapidly,
and higher Landau levels can be occupied,
so the difference of $Y_p$ with and without strong magnetic fields becomes smaller at higher densities.
This feature plays an important role in affecting the chemical potentials.
Compared with the IUFSU model, the TM1 model has a larger symmetry energy slope $L$,
which leads to smaller proton fraction $Y_p$ with the same strength of magnetic fields.
This behavior is consistent with that observed in the case without magnetic fields~\cite{Bao15}.

In Fig.~\ref{fig:4munpe},
we plot the chemical potentials of neutrons, protons,
and electrons as a function of baryon density in pasta phases with $B=0$ and $B=10^{18}$ G.
We can see that the neutron chemical potential $\mu_n$ with $B=10^{18}$ G is smaller than
the results with $B=0$ in all pasta phases,
while the proton chemical potential $\mu_p$ with $B=10^{18}$ G is larger than the one with $B=0$
at lower densities, but $\mu_p$ with $B=10^{18}$ G is smaller than the results with $B=0$
as baryon density $n_b$ increases.
These behaviors can be understood from the features of proton fraction.
The proton fraction with $B=10^{18}$ G is much larger than
the one with $B=0$ at low densities (see Fig.~\ref{fig:3yp}),
so the neutron fraction ($Y_n$) with $B=10^{18}$ G is much lower than the one with $B=0$ accordingly.
As a result, proton (neutron) chemical potential with $B=10^{18}$ G is larger (smaller) than the
results with $B=0$ obviously at low densities,
which also results in the decrease of $\mu_e$ according to the requirement of $\beta$ equilibrium.
Since the difference of proton fraction with $B=10^{18}$ G and $B=0$ becomes smaller at higher densities,
the chemical potentials of neutrons and protons with $B=10^{18}$ G are more close to the results with $B=0$.
The proton chemical potentials with $B=10^{18}$ G are even lower than those with $B=0$
for slab, tube, and bubble phases,
while the neutron chemical potentials with $B=10^{18}$ G and $B=0$ are close to each other.

In Fig.~\ref{fig:5rws},
we show the radii of WS cell $r_{\rm{ws}}$ and
the radii of the inner part of WS cell $r_{\rm{in}}$ with $B=10^{18}$ G and $B=0$
as a function of baryon density $n_b$ for both TM1 and IUFSU models.
It is seen that only four kinds of pasta phases appear in strong magnetic fields $B=10^{18}$ G
for TM1 model, whereas all five pasta phases arise for IUFSU model with or without strong magnetic fields.
One can see that for each solid pasta structure (droplet, rod, and slab),
the radius of WS cell $r_{\rm{ws}}$ decreases with increasing baryon density $n_b$,
but the nucleus radius $r_{\rm{in}}$ increases with $n_b$.
Such feature implies that as baryon density $n_b$ increases,
the size of nucleus becomes larger and the distances between neighboring nuclei become shorter.
In the hollow structure (tube and bubble),
the size of inner gas phase $r_{\rm{in}}$ decreases with $n_b$.
One can see that as baryon density $n_b$ is close to the crust-core transition,
the radius $r_{\rm{ws}}$ increases rapidly,
however, we notice that the binding energy per nucleon is not sensitive to the large $r_{\rm{ws}}$.
The behaviors of $r_{\rm{ws}}$ and $r_{\rm{in}}$ with magnetic fields $B=10^{18}$ G
are similar to the results with $B=0$ as $n_b$ increases.
$r_{\rm{ws}}$ of solid structures (droplet, rod, and slab) with $B=10^{18}$ G is smaller than
that with $B=0$, while $r_{\rm{in}}$ of solid structures with $B=10^{18}$ G is larger.
For tube and bubble phases, $r_{\rm{ws}}$ ($r_{\rm{in}}$) with $B=10^{18}$ G is smaller (larger)
than the results with $B=0$.
As a result, the nuclear radius becomes larger while the separation distance is smaller
with $B=10^{18}$ G compared to the results with $B=0$,
which leads to the volume fraction of dense liquid phase in WS cell increasing more quickly with strong magnetic fields.
Accordingly, the crust-core transition happens at a smaller baryon density $n_b$.
The behavior of $r_{\rm{in}}$ can be understood from the liquid-droplet model.
We know that the competition of Coulomb energy and surface energy plays an important role in determining the
sizes of WS cell and nucleus inside it.
In Ref.~\cite{Lima13}, the authors found that the surface tension increased with the strength of magnetic fields.
A larger surface tension leads to a larger size and more protons of the nucleus inside a WS cell.
As a result, $r_{\rm{in}}$ of droplet, rod, and slab with $B=10^{18}$ G are larger than results of $B=0$.
Furthermore, the radius of WS cell $r_{\rm{ws}}$ also depends on the volume fraction of the inner part,
so its behavior is more complex.

We present in Fig.~\ref{fig:6zad} the charge number $Z_d$ and nucleon number $A_d$
of the spherical nucleus as a function of baryon density $n_b$ in the droplet configuration,
where the background neutron gas is subtracted for defining $A_d$ within the subtraction procedure.
Note that the results of nonspherical configurations are not presented due to arbitrariness in the definition of the nucleus.
It is shown that the charge number $Z_d$ with $B=10^{18}$ G is larger than the one with $B=0$
at fixed baryon density $n_b$.
The reason is that the strong magnetic fields lead to larger proton fraction of WS cell and
larger surface tension, both resulting in more protons in the nucleus.
As baryon density $n_b$ increases, both charge number $Z_d$ and nucleon number $A_d$
increase first and then decrease in the TM1 model with or without strong magnetic fields.
However, the behaviors in the IUFSU model are different,
where both $Z_d$ and $A_d$ increase with increasing baryon density $n_b$.

In order to study further the properties of spherical nucleus of the droplet phase,
we show the distributions of proton $\rho_{p}$, neutron $\rho_n$, and baryon $\rho_b$ in the WS cell
at different average baryon densities $n_b$ in Figs.~\ref{fig:7tmpnb} and \ref{fig:8iupnb} for TM1 and IUFSU model, respectively.
In Fig.~\ref{fig:7tmpnb} (a),
one can see that the proton density $\rho_p$ with $B=0$
decreases with increasing average baryon density $n_b$,
which directly lead to the reduction of $Z_d$ with $n_b$ in Fig.~\ref{fig:6zad} (a),
considering the radius of nucleus $r_{\rm{in}}$ hardly changed at $n_b \leq 0.05$ fm$^{-3}$ for TM1 model (see Fig.~\ref{fig:5rws}).
The behavior of $\rho_p$ with $B=10^{18}$ G is similar to the result with $B=0$,
but the proton with $B=10^{18}$ G has larger range of distribution with increasing $n_b$,
which implies larger nucleus radius.
As a result,
the charge number of nucleus $Z_d$ with $B=10^{18}$ G in Fig.~\ref{fig:6zad} is nonmonotonic for the TM1 model.
At lower density, $n_b=0.01$ fm$^{-3}$, proton density $\rho_p$ with $B=10^{18}$ G
is larger than the one with $B=0$ at fixed radius $r$,
while as $n_b$ increases, $\rho_p$ with $B=10^{18}$ G in the center part of nucleus decreases rapidly
and is lower than the result with $B=0$,
but $\rho_p$ with $B=10^{18}$ G in the outer part of nucleus is always larger than the result with $B=0$.
In general, the nucleus with $B=10^{18}$ G includes more protons compared to the results with $B=0$.
For the same reason, it is easy to understand the behavior of nucleon number $A_d$ in Fig.~\ref{fig:6zad} (b)
from the baryon density distribution in Fig.~\ref{fig:7tmpnb} (c).
Besides, we can see in Fig.~\ref{fig:7tmpnb} that as $n_b$ increases,
the reduction of $\rho_b$ at the center of WS cell comes mainly from the decrease of $\rho_p$,
while the increment of $\rho_b$ at the boundary of WS cell is due to the augment of $\rho_n$ in the gas phase.

By comparing Fig.~\ref{fig:8iupnb} with Fig.~\ref{fig:7tmpnb},
we can see that the effects of strong magnetic fields on nucleon distributions are quite similar in IUFSU
and TM1 models.
The presence of strong magnetic fields can enhance the charge number in the nucleus and reduce the neutron density $\rho_n$ and baryon density $\rho_b$ both at the center and boundary of WS cell.
It is shown that the nucleon distributions in WS cell for IUFSU model are different from the results for TM1 model.
From Fig.~\ref{fig:8iupnb}, we can see that
$\rho_p$ in the center of nucleus decreases more quickly with increasing $n_b$ than the results for TM1 model,
especially the results with strong magnetic fields $B=10^{18}$ G,
which decreases about 50$\%$ from $n_b=0.01$ to $0.03$ fm$^{-3}$.
The increment of charge number in nucleus $Z_d$ with increasing $n_b$
is due to the increase of nuclear radius $r_{\rm{in}}$ and larger $\rho_p$ in the boundary area of nucleus.

In order to investigate the effects of strong magnetic fields on the distributions of nucleons and leptons of various pasta phases,
we show in Fig.~\ref{fig:9npe}
the density distributions of neutrons, protons, and electrons in WS cell at five different average baryon densities $n_b=0.03,\ 0.05,\ 0.07,\ 0.08,\, 0.09$ fm$^{-3}$ for IUFSU model with $B=0$ and $10^{18}$ G, respectively.
From Fig.~\ref{fig:9npe} (a), we can see that the neutron density $\rho_n$ at the center of the WS cell is larger than that at the boundary for droplet, rod, and slab phases,
while it is opposite for the tube and bubble phases.
We also notice that the difference of $\rho_n$ between the center and the boundary of the WS cell  decreases with increasing $n_b$,
which implies that nuclear distribution in the WS cell becomes more diffuse as close to the crust-core transition.
A similar tendency is also observed in Fig.~\ref{fig:9npe} (c),
where the electron distributions in the WS cell are plotted.
We can see that the electron density $\rho_e$ is close to uniform distribution with increasing $n_b$.
With the strong magnetic field $B=10^{18}$ G, the electron density in the whole WS cell obviously increases
comparing to the results with $B=0$.
This is different from the effects of strong magnetic fields on proton distribution $\rho_p$.
From Fig.~\ref{fig:9npe} (b),
we can see that $\rho_p$ at the boundary of nucleus with $B=10^{18}$ G is
larger than the results with $B=0$.
However, $\rho_p$ in the center of nucleus with $B=10^{18}$ G is lower than the one with $B=0$.
We can clearly see that the proton disappears in the gas phase due to its chemical potential smaller than its mass.

Note that some kinks in $\rho_p$ with $B=10^{18}$ G correspond to the changes of Landau level.
For clarity, the neutron density in the center of nuclear liquid phase $n_{n,L}$,
the one in the gas phase $n_{n,G}$, and the proton density in the center of  nuclear liquid phase $n_{p,L}$ are plotted in Fig.~\ref{fig:10nnnp} as a function of baryon density $n_b$
with $B=0$ and $B=10^{18}$ G for IUFSU model.
One can see that as $n_b$ increases,
the neutron density of liquid phase $n_{n,\,L}$ does not change much;
however, the neutron density of gas phase $n_{n,\,G}$ increases with $n_b$ obviously.
Comparing to the results with $B=0$, both $n_{n,\,L}$ and $n_{n,\,G}$ with $B=10^{18}$ G decrease.
On the other hand,
the proton density of liquid phase $n_{p,L}$ decreases with increasing $n_b$.
At lower baryon densities, such as in the droplet phase, $n_{p,\,L}$ with $B=10^{18}$ G is higher than that with $B=0$,
while at higher baryon densities, the behavior is opposite.

\section{Conclusion}

\label{sec:4}

In this work, we have studied the influence of strong magnetic fields on the properties
of nuclear pasta phases and crust-core transition in the inner crust of neutron star
by using the RMF model and the self-consistent TF approximation.
The distributions of nucleons and electrons in the WS cell are determined self-consistently,
in which the charge neutrality and $\beta$ equilibrium conditions are satisfied.
It has been found that the pasta phase structures and the crust-core transition density
were changed obviously when the magnetic field strength is as large as $B=10^{18}$ G,
where the binding energy per nucleon $E/N$ is lower than the results with $B=0$,
and the onset densities of various pasta phases and crust-core transition density become smaller.
However, the proton fraction $Y_p$ with $B=10^{18}$ G is larger than that with $B=0$,
since the protons occupy the lowest Landau level.
The impacts of anomalous magnetic moments of nucleons are almost invisible in the case of $B=10^{17}$ G,
but they have to be taken into account for a stronger magnetic field as $B=10^{18}$ G.
In general, the radius of WS cell decreases with increasing $B$,
while the size of nucleus increases with $B$,
which results in the charge number and nucleon number of the nucleus varying with $B$.
The density distributions of nucleons and electrons with $B=10^{18}$ G
are clearly different from the results with $B=0$.

In order to check the model dependence of the results obtained,
we adopt two successful RMF models, i.e., TM1 and IUFSU, with different symmetry energies and their slopes,
which play an important role in determining
the properties of inner crust of neutron star with strong magnetic fields.
The features with strong magnetic fields due to the symmetry energy and its density slope
are similar to the results with $B=0$,
which are consistent with our earlier study~\cite{Bao15}.
A smaller slope $L$ leads to more complex pasta structures.
For the TM1 model with a larger slope $L$,
only droplet appears in the inner crust of neutron star for $B=0$.
However, some nonspherical pasta phases arise before crust-core transition for $B=10^{18}$ G,
even though the crust-core transition density becomes smaller.
It would be interesting to further study the nuclear pasta phase with strong magnetic fields
and their impacts on the observations of neutron star.


\section*{Acknowledgment}

This work was supported in part by the National Natural Science Foundation
of China (Grants No. 11805115, No. 11675083, and No. 11775119).


\end{document}